\documentclass[graybox,natbib,nosecnum]{svmult}
\bibpunct{(}{)}{;}{a}{}{,} 

\pdfoutput=1   

\usepackage{mathptmx}       
\usepackage{helvet}         
\usepackage{courier}        
\usepackage{type1cm}        
\usepackage{amssymb}                          

\usepackage{makeidx}         
\usepackage{graphicx}        
\usepackage{multicol}        
\usepackage[bottom]{footmisc}
\usepackage[normalem]{ulem}	
\usepackage{hyperref}  

\usepackage{soul}   

\newcommand{\hbindex}{}

\makeindex             

\begin{document}

\title*{Dust evolution in protoplanetary disks}
\author{Nienke van der Marel  and Paola Pinilla}
\institute{Nienke van der Marel \at Leiden Observatory, Niels Bohrweg 2, 2333 CA Leiden, The Netherlands, \email{nmarel@strw.leidenuniv.nl}
\and Paola Pinilla \at Mullard Space Science Laboratory, University College London, Holmbury St Mary, Dorking, Surrey, RH5 6NT, UK \email{p.Pinilla@ucl.ac.uk}}
%
%
\maketitle

\abstract{
Planet formation models rely on knowledge of the physical conditions and evolutionary processes in protoplanetary disks, in particular the grain size distribution and dust growth timescales. In theoretical models, several barriers exist that prevent grain growth to pebble sizes and beyond, such as the radial drift and fragmentation. Pressure bumps have been proposed to overcome such barriers. In the past decade ALMA has revealed observational evidence for the existence of such pressure bumps in the form of dust traps, such as dust rings, gaps, cavities and crescents through high-resolution millimeter continuum data originating from thermal dust emission of pebble-sized dust grains. These substructures may be related to young protoplanets, either as the starting point or the consequence of early planet formation. Furthermore, disk dust masses have been measured for complete samples of young stars in clusters, which provide initial conditions for the solid mass budget available for planet formation. However, observational biases exist in the selection of high-resolution ALMA observations and uncertainties exist in the derivation of the disk dust mass, which both may affect the observed trends. This chapter describes the latest insights in dust evolution and disk continuum observations. Specifically, disk populations and evolutionary trends are described, as well as the uncertainties therein, and compared with exoplanet demographics.}


\section{Introduction}
\begin{figure}
    \centering
    \includegraphics[width=\textwidth]{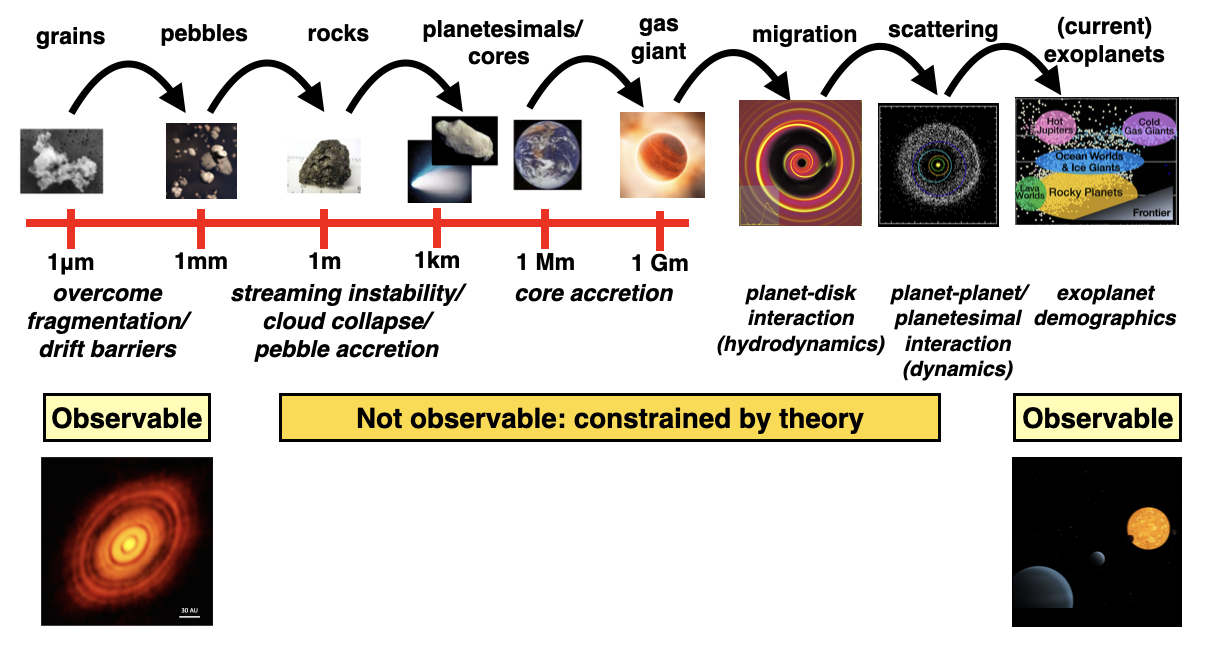}
    \caption{Schematic of the growth from micron-sized grains to cores and planets}
    \label{fig:planetformation}
\end{figure}

Planets form in protoplanetary disks of gas and dust in the early stages of the disk, during or right after the formation of the protostar. The formation scenario is generally visualized as the \hbindex{coagulation} of small dust grains into larger and larger particles through collisions. This process involves the growth from sub-micron sized dust particles up to 1000s of kilometer sized planets and planet cores, or more than 12 orders of magnitude in size and more than 36 orders of magnitude in mass within less than 10 Myr (Figure \ref{fig:planetformation}). The formation of the core may be followed by gas accretion, creating a gas giant, which must happen before the gas disk has dissipated. The mean gas \hbindex{disk lifetime} has long been thought to be 2-3 Myr \citep{Hernandez2007}, but recent work suggests that the mean lifetime may be as long as 5-10 Myr \citep{Michel2021,Pfalzner2022}. The key issue in understanding the planet formation process is thus primarily understanding the formation timescales of rocky cores through dust growth.

Various disk processes are responsible for both aid and hindrance of the efficiency of this coagulation, where the combined efficiencies and timescales of these processes have not been fully understood yet. One of the major complexities in this problem is that in protoplanetary disks only the smallest dust pebbles (micron-sized up to centimeter-pebble-sized) are observable, as larger boulders and planetesimals do not emit efficiently any more. Only the end product, the exoplanet population, becomes observable again, but all the steps in between remain mostly constrained by formation models, and by clues from remnants of the planet formation process in the Solar System. However, protoplanetary disk observations can provide answers as well through statistical studies and spatially resolved observations of substructures, which may show indirect evidence for early planet formation, as well as provide constraints on the initial conditions of the birth environments of planets. 

The research field of protoplanetary disks has evolved tremendously in the past two decades, through new observational facilities in infrared and (sub)millimeter and developments in disk dynamics and dust evolution modeling. In particular the Atacama Large Millimeter/submillimeter Array (\hbindex{ALMA}) and optical/infrared (OIR) instruments at 8m-class telescopes like VLT, Gemini and Subaru have provided a wealth of data in the form of spatially resolved images of dust and gas in disks, both in the early embedded stage where the disk is still surrounded by an envelope (also called Class 0 or Class I stage), the massive protoplanetary disk stage (Class II) and the evolved debris disk stage consisting of stars surrounded by Kuiper-belt equivalents. This chapter aims to summarize to most important insights in the study of dust in protoplanetary disks from the past decade and their implications for exoplanets and planet formation, as well as the open questions and issues that remain under debate. Observational constraints on the disk gas composition are described in the chapter `Chemistry During the Gas-Rich Stage of Planet Formation'.

This chapter is organized as follows. First, the main dust evolution processes are reviewed, including radial drift, trapping and transport, as well as challenges in dust evolution models. Second, an overview of observations of large-scale substructures in protoplanetary disks is presented, with most emphasis on (sub)millimeter observations with ALMA but some discussions on results from OIR imaging and VLTI results as well. In particular, the different types of substructures are discussed, including observational evidence for their role in the evolution of the disk and the biases in the detectability of substructures so far. Third, the main methods to derive disk dust masses and the uncertainties in this approach are reviewed, as well as the consequences for the solid mass budget available for planet formation. Observational constraints on grain sizes in disks are discussed as well. A discussion on trends in exoplanets with stellar mass and metallicity and possible similarities in disk demographic studies is included. The chapter finishes with a summary and an outlook to future facilities. 


\section{Dust Evolution in Protoplanetary Disks}

Protoplanetary disks are the sub-product of angular momentum conservation in the star formation process, where the angular momentum of molecular clouds is much higher than the resulting protostars. As a consequence, the angular momentum is re-distributed in a rotating disk around the young star. In these disks, which are the birth sites of planets, the angular momentum and mass are transported by different physical mechanisms, such as the \hbindex{magneto-rotational instability} \citep[MRI,][]{pringle1981}, or a \hbindex{magnetohydrodynamical wind} \citep[e.g.,][]{bai2013}, while the material is accreted onto the star.

The initial abundances of protoplanetary disks are thought to be similar to the interstellar medium, which means that most of the material is molecular gas ($\sim$99\% of the mass), while only $\sim$1\% is dust with an initial size of (sub-)micron-sized particles. These particles are the seeds for the formation of pebbles and planetesimals, which are the building blocks of planets. 

The growth from interstellar dust particles to pebbles is a complex process that is regulated by the interaction of the dust with the gas. In this process, particles move in the disk due to different velocities, including: Brownian motion, \hbindex{settling}, \hbindex{radial drift}, and \hbindex{turbulence} \citep[e.g.,][]{brauer2008, birnstiel2010}. How relevant each of these velocities are to the total velocity of particles depends on how well the particles are coupled to the gas, which is measured by the  importance of the drag
forces by comparing stopping time of the particles to the Keplerian frequency. This is quantified by the \hbindex{Stokes number}, which is the stopping time normalized by the orbital timescale. While different drag regimes exist, dust particles in the gaseous disk typically fall in the Epstein drag regime, where the particle size is smaller than the mean free path of gas molecules. In the Epstein regime, the Stokes number is defined as

\begin{equation}
    St= \frac{a \rho_s}{\Sigma_g}\frac{\pi}{2},
    \label{eq:Stokes}
\end{equation}

\noindent where $a$ is the grain size, $\rho_s$ is the intrinsic volume density of the particles, and $\Sigma_g$ is the gas surface density. Equation \ref{eq:Stokes} is valid near or at the midplane of the disk. For particles with $St\ll$1 the main source for the relative velocities is Brownian motion and settling, and these are usually lower than the threshold velocity for catastrophic collisions. Hence, these dust motions lead to \hbindex{grain growth} \citep{Birnstiel2016}. When the Stokes number of particles increases and reach values near unity, the turbulent and drift motions become the dominant sources for their velocities, which can lead to destructive collisions or the loss of dust particles towards the star in timescales that are shorter than few thousands of years. 
The threshold velocity for catastrophic collisions or \hbindex{fragmentation velocity} is one of the main unknown parameters in models of dust evolution. The models rely on laboratory experiments and numerical simulations for limits on this quantity. The fragmentation velocity depends on surface energy of the grains and therefore on the particle composition \citep[e.g.,][]{blum2000, wada2011, musiolik2016}. Although water-ice particles are expected to be more sticky than silicates, recent laboratory experiments that reach conditions similar to those in the Universe (low density, pressure and temperature) have challenged this idea \citep{Gundlach2018, musiolik2019, steinpilz2019}. Currently, it is thought that the fragmentation velocity of particles has a value from few to tens of \,m\,s$^{-1}$.

The growth process to sub-micron sized particles to pebbles is accessible observationally, since different wavelengths can trace different dust size particles. For example, scattered light observations at the optical and near infrared trace the distribution of micron-sized particles located in the upper layers of the disk. On the other hand, pebbles which are located in the midplane of the disk can absorb the light of similar wavelength and re-emit the light. This process can also be affected by scattering, as discussed later on.
However, observationally accessing the information of boulders and planetesimals in protoplanetary disks is very challenging because of their very low opacities. As a result, there is an observational gap between pebbles and planets, that  can only be closed by connecting dust/planetesimal evolution models to exoplanets statistics. 

The following sections present the most important processes of the first steps of planet formation, in particular radial drift and its connection with observations. In this topic, a brief overview of the origin of substructures in disks is presented. This section ends with the current challenges in this field. 

\subsection{Dust Radial Drift} \label{sect:drift}

Depending on their \hbindex{Stokes number}, particles experiences different velocities. One of the main source for dust velocities with particles with Stokes number near unity is \hbindex{radial drift}. The drift velocity of particles originate from the sub-Keplerian  azimuthal gas velocity in a disk where pressure decreases monotonically with radius. The total drift velocity of dust particles in protoplanetary disks is given by \citep{Weidenschilling1977}:

\begin{equation}
        v_{\mathrm{drift}}(\mathrm{St}, M_\star, L_\star)=\frac{1}{\textrm{St}^{-1}+\textrm{St}} \frac{\partial_r P (M_\star, L_\star)}{\rho(M_\star, L_\star) \Omega(M_\star)}. 
        \label{eq:vdrift} 
\end{equation}

There is a dependency of $v_{\mathrm{drift}}$ on the stellar mass and luminosity that makes radial drift to be more efficient around low-mass stars and for stars more massive than $>2.5$\,$M_\odot$ \citep{Pinilla2013, Pinilla2022, zhu2018}.

In Eq.~\ref{eq:vdrift} $\rho$ is the disk gas volume density, which in hydrostatic balance and in a vertically isothermal disk is given by

\begin{equation}
        \rho=\rho_0(M_\star, L_\star)\exp\left(\frac{-z^2}{2h^2(M_\star, L_\star)}\right),
        \label{eq:rho}
\end{equation}

\noindent where $\rho_0$ is the midplane density, $h$ is the disk scale height, and $P$ is the isothermal pressure, each of them given by

\begin{eqnarray}
        &\rho_0&=\frac{\Sigma_g(M_\star)}{\sqrt{2\pi}h(M_\star, L_\star)}\\
        &h(M_\star, L_\star)&=\frac{c_s(L_\star)}{\Omega(M_\star)}\\
        &P&= \rho(M_\star, L_\star) c_s^2(L_\star),
        \label{eq:rho_0}
\end{eqnarray}

\noindent respectively. The assumption that $\Sigma_g$ depends on the stellar mass comes from assuming that the total disk mass scales with the stellar mass, supported by observations as described later.  The sound speed ($c_s$) is given by:

\begin{equation}
   c_s^2= \frac{\sigma_{\textrm{SB}} T_{\textrm{disk}} (L_\star)}{\mu m_p},
        \label{eq:cs}
\end{equation}

\noindent with $\sigma_{\rm{SB}}$, $\mu,$ and $m_p$ the Stefan-Boltzmann constant, the mean molecular mass, and the proton mass, respectively. When assuming a \hbindex{disk temperature} ($T_{\rm{disk}}$) of a passive disk, there is a  dependency between the disk temperature and stellar luminosity as  \citep{Kenyon1987}:

\begin{equation}
        T(r, L_\star)=T_\star\left(\frac{R_\star}{r}\right)^{1/2} \phi_{\rm{inc}}^{1/4} =\left(\frac{L_\star \phi_{\rm{inc}}}{4\pi \sigma_{\textrm{SB}} r^2}\right)^{\frac{1}{4}}
  \label{eq:temp}
,\end{equation}

\noindent where $R_\star$ and $T_\star$ are the stellar radius and temperature, respectively, and  $\phi_{\rm{inc}}$ is the incident angle, which usually is a very small quantity.

\begin{figure*} 
    \centering
    \tabcolsep=0.05cm 
    \begin{tabular}{cc}   
        \includegraphics[width=0.5\columnwidth]{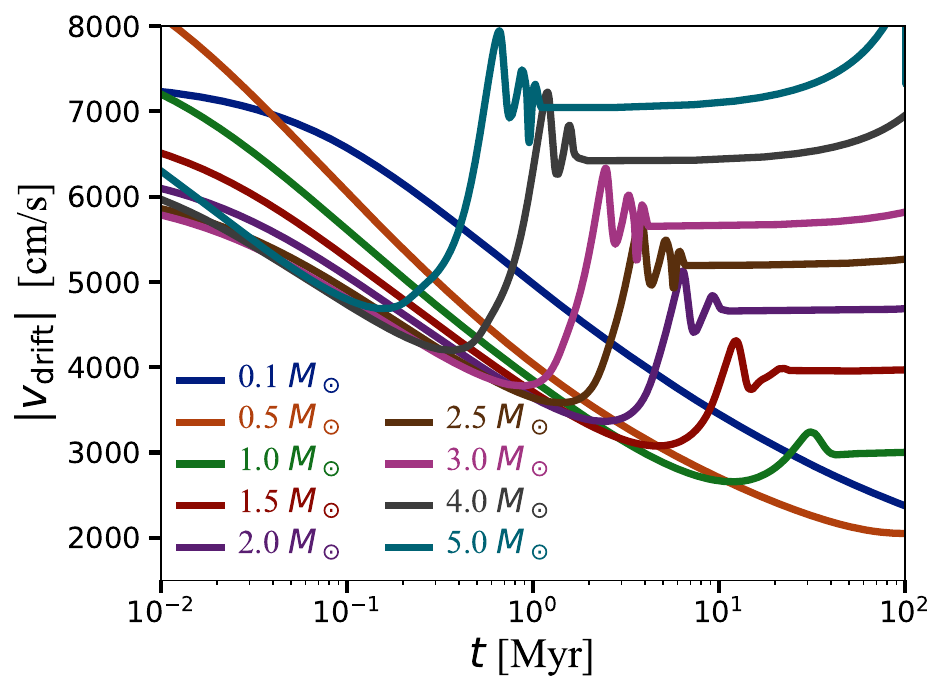}&
        \includegraphics[width=0.5\columnwidth]{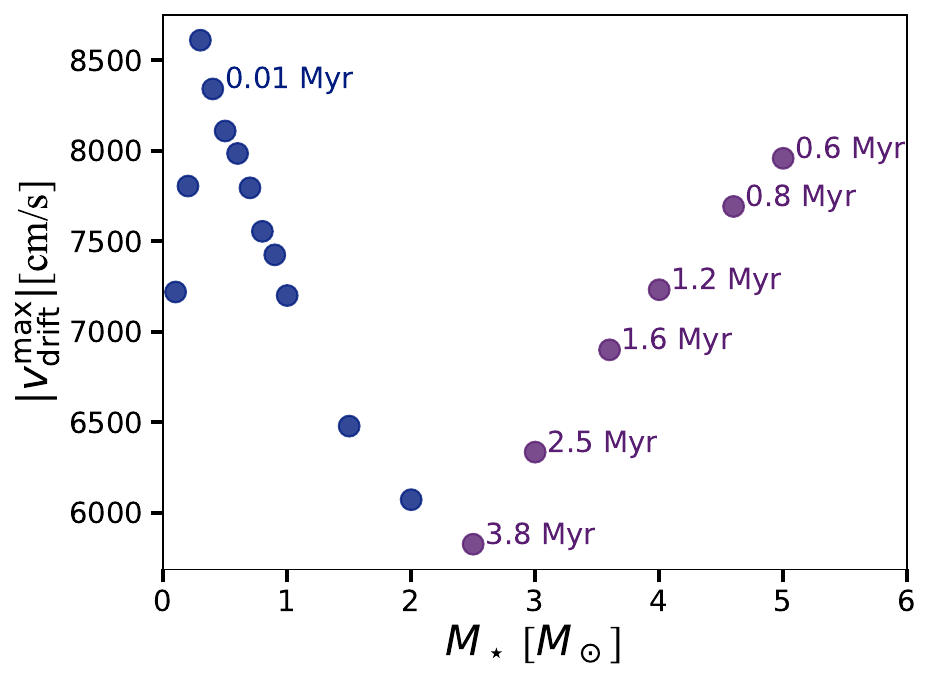}
    \end{tabular}
    \caption{\emph{Left:} absolute value of the drift velocity when St=1 as a function of time around stars with different masses, when assuming evolutionary tracks from \citet{dotter2008} up to 30\,Myr of 0.1-5\,$M_\odot$ stars in the H-R diagram \citep[adapted from][]{Pinilla2022}. \emph{Right:} absolute value of the drift velocity when St=1 as a function of stellar mass. The values are taken at 0.01\,Myr for all the stars with a mass $<2.5$\,$M_\odot$, while for more massive stars the drift velocity value is taken when it is maximum and this time is indicated for each point.}
    \label{fig:drift}
\end{figure*}

These dependencies on stellar mass and luminosity, leaves to $v_{\mathrm{drift}}\propto L_\star^{1/4}/\sqrt{M_\star}$.  The left panel of Fig.~\ref{fig:drift} shows the absolute value of the drift velocity when St=1 as a function of time around stars with different masses, when assuming evolutionary tracks from \cite{dotter2008} up to 30\,Myr of 0.1-5\,$M_\odot$ stars. The drift velocities for particles with St=1 are pretty high, between $\sim5000-8000$cm~s$^{-1}$, implying that grains with such a Stokes number will be quickly lost towards the star, for example at one astronomical unit, they will drift in less than 100\,yr. This is the origin of the well-known \emph{\hbindex{radial drift barrier}}, which is due to high radial drift velocities that can lead to the lost of particles onto the star, or to the fragmentation of grains when they collide at these high drift velocities.

The drift barrier occurs when growth is limited by radial drift and is given by 

\begin{equation}
	a_{\mathrm{drift}}=\frac{2 \Sigma_d}{\pi\rho_s}\frac{v_K^2}{c_s^2}\left \vert \frac{d \ln P}{d\ln r} \right \vert^{-1},
  \label{eq:adrift}
\end{equation}

\hbindex{Fragmentation} due to turbulent velocities also can limit the growth of dust particles. \hbindex{Turbulence} induces dust motion that depends on the dust particle size, this motion for example diffuse particles from dust particle concentrations as discussed later. Assuming that the disk turbulence is described by an effective \hbindex{viscosity}, the well-known $\alpha$-parameter \citep{Shakura1973}, and assuming that particles stick in collisions below a given fragmentation speed $v_{\rm{frag}}$, the fragmentation limit is 

\begin{equation}
	a_{\mathrm{frag}}=\frac{2}{3\pi}\frac{\Sigma_g}{\rho_s \alpha}\frac{v_{\rm{frag}}^2}{c_s^2}.
  \label{eq:afrag}
\end{equation}

The disk viscosity parameter $\alpha$ remains one of the most unknown quantities in models of planet formation. It is usually assumed constant throughout the disk. However, its value depends on the dust properties and its abundance, the magnetic field strength and its configuration, the gas and dust chemistry, and the various ionization sources \citep{Delage2022, Lesur2022}, whereas many of these properties cannot be constrained from disk observations. Theoretical models predict that when the disk is now well ionized, the disk viscosity can be low, which usually happens in the dense inner regions at the midplane. This region, known as the \hbindex{dead zone}, is expected to have a lower effective viscosity than the rest of the disk.

Using typical disk and stellar parameters, the \hbindex{maximum grain size} is set by \hbindex{fragmentation} in the first $\sim$20\,au \citep[e.g.][Figure 4]{Birnstiel2016}, with values between few centimetres to few millimetres, while in the outer disk is limited by drift with values from few millimetres to hundreds of microns. 

\subsubsection{Dependency of radial drift with stellar mass}

For different stellar masses, the maximum drift velocity in the left panel of Fig.~\ref{fig:drift} is taken and plotted against the stellar mass in the right panel of Fig.~\ref{fig:drift}. For stars with a mass $<2.5$\,$M_\odot$, this value is taken at early times (0.01\,Myr) when drift has its highest values, and for more massive stars, the time when the drift velocity is maximum, which is displayed. 
This figure shows the binomial behaviour of the \hbindex{radial drift}. Specifically, for low-mass stars (0.1-0.5\,$M_\odot$) the drift velocity is higher than for a Solar-mass star throughout the disk lifetime. While for stars more massive than $2.5$\,$M_\odot$, the drift velocities are lower at early times and sharply increase when the stellar luminosity also increases and this happens within the disk lifetime ($<$5-10\,Myr).

This interesting dependency of radial drift on stellar properties may explain different observationally aspects of protoplanetary disks around different stellar masses. An example is the lack of disk detections around disks around stellar objects more massive than 2.5\,$M_\odot$ at ages older than 3$-$4 Myr \citep{Benisty2022}. In addition, this suggests that if pebbles are present in disks around low-mass stars, whichever physical process halting the drift should be more efficient around such stellar objects \citep[][]{Pinilla2013}.

\subsection{Solution to the Radial Drift Barrier and Origins of Substructures}

\begin{figure*} 
    \centering
    \includegraphics[width=\columnwidth]{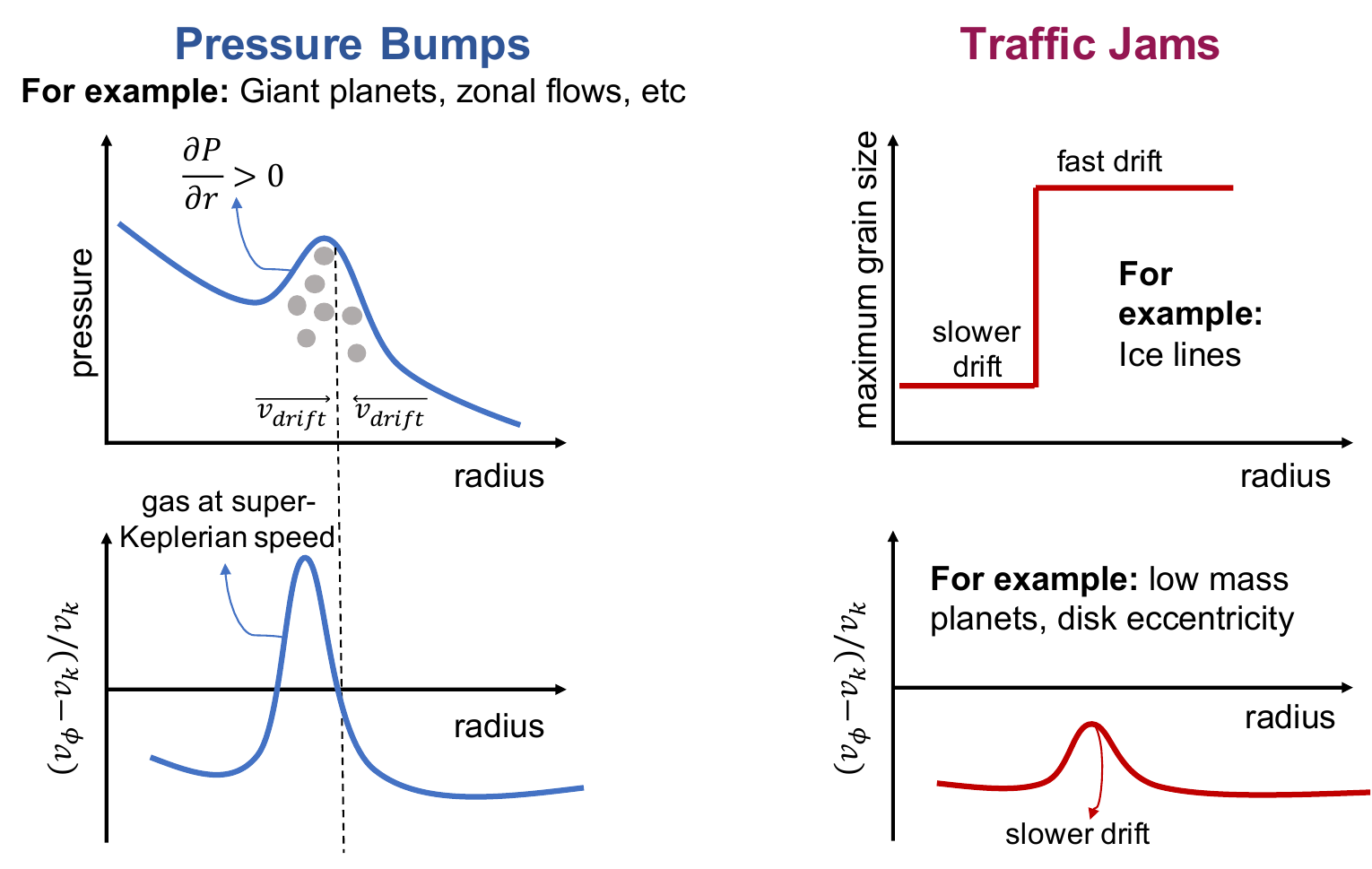}
    \caption{Sketch to compare pressure bumps vs. traffic jams. In pressure bumps, the gas azimuthal velocity is positive where the pressure gradient is positive and as a result dust particles drift outwards, while they drift inwards in the rest of the disk where the pressure gradient is negative (Eq.~\ref{eq:adrift}). Traffic jams have different origins in disks. For example, when there are variations of the maximum grains size as expected near ice-lines of the main volatiles in disks. In the locations where the particle size is higher, particles drift faster due to their higher Stokes number and vice versa. Other possibilities of traffic jams happen when there is a perturbation in azimuthal gas velocity that makes particles to slow down their drift without the gas moving super-Keplerian (i.e, without a pressure bump).}
    \label{fig:bumps_jams}
\end{figure*}

There are two main solutions to reduce or totally stop the radial drift of particles in protoplanetary disks. One is to increase the cross-section of the particles, such that their Stokes number is low while they are still large \citep[e.g.,][]{kataoka2013}. This will imply very fluffy aggregates that can be  compressed by gas pressure and/or self-gravity of the aggregates. In this scenario, the compression processes need to be fast enough to avoid the drift barrier. The other alternative that has gained a lot of attention in the last decade is the presence of pressure bumps or traffic jams in protoplanetary disks.

In \hbindex{pressure bumps}, there is a region where the pressure gradient is positive and as a consequence the particles drift outwards (Eq.~\ref{eq:vdrift}). In the pressure maximum, the drift velocity of particles is zero. Dust evolution models of particle traps or \hbindex{dust traps} have shown that the efficiency of trapping depends on the dust diffusion (or disk viscosity) and the pressure gradient. In a pressure bump, the disk viscosity sets the \hbindex{maximum grain size}, and hence the Stokes number of the particles that are trapped. Particles are efficiently trapped when St$\gtrsim \alpha$, otherwise \hbindex{radial diffusion} makes particles escape the bump and trapping is inefficient \citep[e.g.,][]{Pinilla2012, ovelar2016}.

Some examples of particle traps are when a planet (usually more massive than Neptune) opens a \hbindex{planetary gap} in the gas surface density that results in a pressure bump at the outer edge of the gap \citep[e.g.,][]{Zhu2011, Gonzalez2012, Dong2015gaps, Zhang2018}; variations of the disk magnetic field that leads to overdensities in the disk that result in pressure bumps as well \citep[the well-known \hbindex{zonal flows},][]{Uribe2011, Flock2015}; or at the edge of a cavity formed by \hbindex{photoevaporation} by the central star \citep[e.g.,][]{Alexander2006, Owen2010, Picogna2019}. In the recent review by \cite{Bae2022}, there is a extensive discussion about the potential origins of pressure bumps in disks.

In \hbindex{traffic jams}, the radial (or azimuthal) drift of particles is reduced by different mechanisms, without the presence of pressure bumps.  For example, it is thought that near \hbindex{ice-lines} of main volatiles, particles change their composition and hence their capability to stick. For instance, inside the water ice-line where grains are mostly silicates, fragmentation due to collisions is efficient in these regions. As a consequence, the \hbindex{maximum grain size} is reduced (Eq.~\ref{eq:afrag} when $v_{\rm{frag}}$ is lower). As a consequence of the maximum grain size reduction, particles experience lower radial drift due to their low Stokes number (Eq.~\ref{eq:vdrift} and Eq.~\ref{eq:Stokes}). How the maximum fragmentation velocity changes for different grain composition is still an open question. \cite{Pinilla2017} performed numerical simulation of dust evolution in the presence of one, two, or three ice-lines affecting the dust growth and dynamics, and produced observational predictions of the gaps and rings in disks.

Other alternatives for dust traffic jams in disks are when a low-mass planet is embedded in the disk or when the disk is eccentric. In the first case, a low mass planet that does not open a gap would nevertheless affect the azimuthal velocity of the gas, decreasing the drift velocity near the planet location \citep[Fig.~\ref{fig:bumps_jams}, e.g.,][]{Rosotti2016}. These traffic jams can create also multiple rings and gaps that are observable with telescopes like ALMA.

Pressure bumps and traffic jams produced by different physical mechanisms can explain the structures observed in protoplanetary disks. Figure~\ref{fig:all_cases} shows examples of dust density distributions after $\sim 1-1.5$\,Myr of evolution of dust evolution models performed with \emph{Dustpy} \citep{Stammler2022}. These models include the dynamics and different growth processes (sticking, fragmentation, and erosion) of particles. The results from Figure~\ref{fig:all_cases} show:

\begin{figure*} 
    \centering
    \includegraphics[width=\columnwidth]{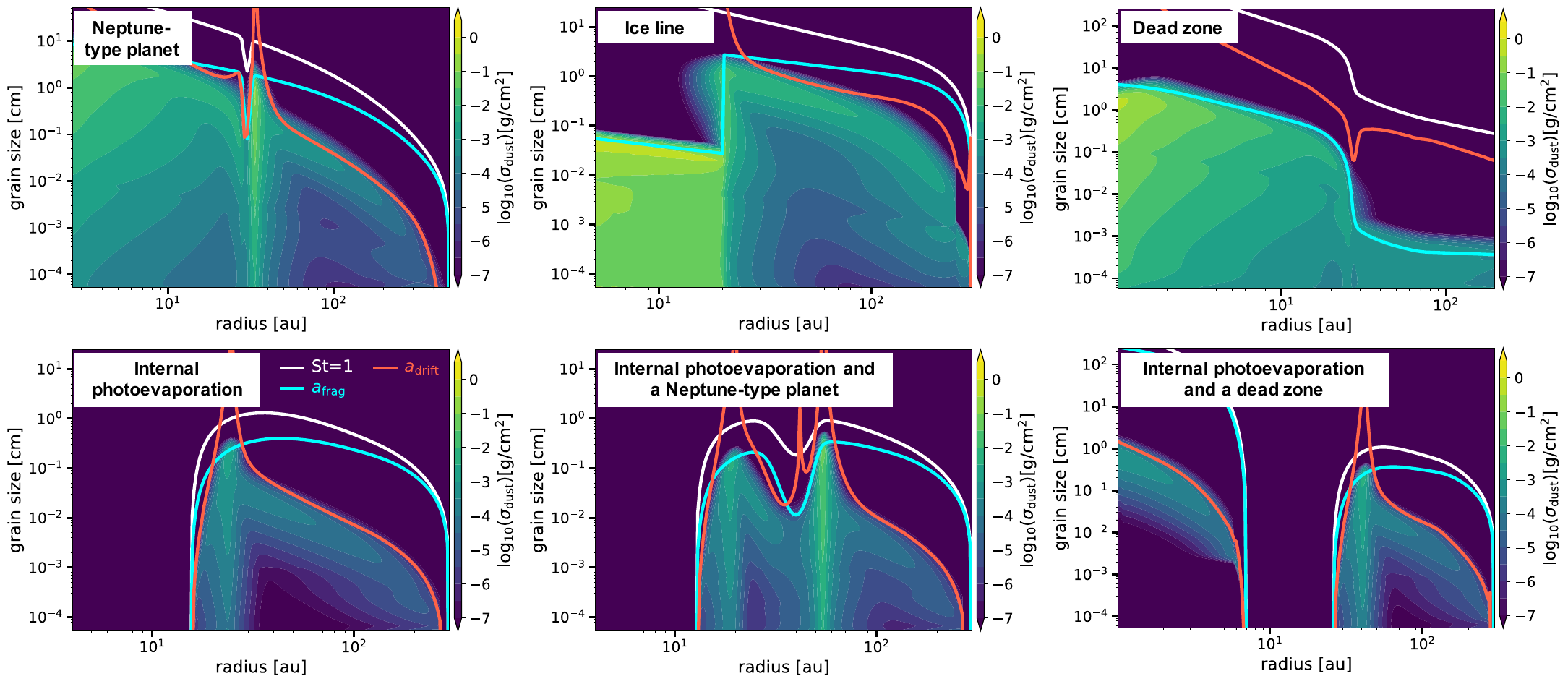}
    \caption{Dust density distribution from dust evolution models after $\sim 1-1.5$\,Myr of evolution for all the cases. From left-top panel are the models of: a Neptune-type planet (pressure bump); an ice-line (traffic jam) artificially set at 20\,au where the fragmentation velocity changes from  1 inside to 10\,m\,s$^{-1}$ outside 20\,au \citep{Garate2020}; dead zone \citep[traffic jam, models from][]{delage2023}; internal photoevaporation (pressure bump) models from Garate el al. (submitted); and combinations of models of internal photoevaporation with either a Neptune-type planet or a dead zone \cite{Garate2021} and submitted. The solid white line represents a Stokes number of unity (which is proportional to the gas surface density), while cyan and red show the fragmentation (Eq.~\ref{eq:afrag}) and drift (Eq.~\ref{eq:adrift})barrier, respectively.}
    \label{fig:all_cases}
\end{figure*}

\begin{itemize}
    \item A Neptune-type planet embedded in the disk around 30\,au. In this case, it is possible to see the dust trapping that the edge of the planetary gap. The more massive the planet is, a deeper and wider gap is created when assuming the same disk viscosity and thermodynamical properties. When the planet reaches a mass of $\sim$1\,$M_{\rm{Jup}}$, it can open a very deep and wide gap in the distribution of pebbles, resembling properties of so-called \hbindex{transition disks} \citep[e.g.,][]{Zhu2011, Pinilla2012}.
    \item An artificial ice-line located at 20\,au. In this simulation, the fragmentation velocity changes at 20\,au from  1 inside to 10\,m\,s$^{-1}$ outside 20\,au. Because  dust particles reduce their size and their Stokes number near this location, they slow down their drift. These density distributions can lead to clear substructures in the disk \citep{Banzatti2015, Pinilla2017, Garate2020}.
    \item A \hbindex{dead zone} which extends up to $\sim$20\,au. In this case there are variations of the disk viscosity that allow dust particles to grow larger inside 20\,au (opposite to the case of the ice-line). Dust particles in the outer disk experience a reduction of their radial drift. Models are taken from \cite{delage2023}, where there is no self-consistent evolution of gas and dust yet, hence there is no trapping, but a traffic jam effect. 
    \item Internal \hbindex{photoevaporation} alone or in combination with an Neptune-type planet or a dead zone. Photoevaporation by the irradiation of the central star can open a very deep gap and particles are trapped at the edge of that gap. If a Neptune-type planet is embedded in the disk, there can be multiple regions of trapping at Myr timescales (Garate et al., submitted). In the case of internal photoevaporation combined with a dead zone, a wide gap is formed in the distribution of pebbles, while keeping an inner disk and averaged accretion rates as observed in transition disks \citep{Garate2021}.
\end{itemize}

Although \hbindex{planet-disk interaction} remains as one of the most used scenarios to explain observed substructures (see next section), other scenarios cannot be excluded until the planets are confirmed. 

\subsection{Dust Settling in Disks}
From micron to sub-millimeter sized dust grains, \hbindex{settling} is an import source for the velocity of the dust and their collisions. The settling velocity of grains  is given by the balance of  the gas drag force, vertical gravitational force, and the vertical stirring (or vertical turbulence) of particles. Large particles settle faster to the midplane due to gravity, which are also less affected by vertical stirring.  The particle scale height $h_{\mathrm{d}}$ for a grain with a given Stokes number is \citep{youdin2007, birnstiel2010}

\begin{equation}
	h_{\mathrm{d}}(St)=h \times \mathrm{min} \left( 1,\sqrt{\frac{\alpha}{\mathrm{min}(St,1/2)(1+St^2)}}\,\right).
\label{eq:dust_scaleheight}
\end{equation}

Comparing optical and near-infrared observations  that trace the distribution of the micron-sized particles and millimeter observations that trace the large particles that are subject to dust settling, \citet{Villenave2020, Villenave2022, Villenave2023} provided constraints on the level of vertical turbulence, which in most cases is quite moderate, with $\alpha$ as low as $10^{-5}-10^{-4}$. Furthermore, high degrees of settling can enhance \emph{grain growth} due to the locally high dust concentration.

\subsection{Current challenges in dust evolution models }

There are several physical processes that can affect dust evolution and that remain unconstrained from observations. For example, the disk temperature is a fundamental quantity of disks, and it can affect different properties, such as the disk composition and  locations of ice lines of main volatiles in protoplanetary disks, hence where structures potentially formed by ice-lines are expected. 

The fragmentation velocity for dust particles for different dust compositions remain still an open question. Most of the laboratory experiments of dust collisions do not have yet the temperatures and pressures that are expecting in the disk midplane, and hence it remains unknown the sticky properties of particles with different compositions. This parameter has a large impact in the maximum grain size that particles can reach through collisional growth (Eq.~\ref{eq:afrag})

In addition, the disk \hbindex{viscosity} and turbulence are generally unknown in disks, and it directly affects the dust settling and growth. From observations of edge-on disks, the comparison of the distribution of micron-sized and millimeter-sized particles can provide insights about the vertical disk turbulence (Eq.~\ref{eq:dust_scaleheight}), which in most cases suggest that disk turbulence is low. Other alternatives to obtain the disk turbulence is to measure the non-thermal broadening of molecular lines due to turbulence \citep[e.g.,][]{Flaherty2015, Flaherty2020}, and to measure the radial width of the sub-structures in disks \citep[which are expected to be very narrow in the absence of turbulence that diffuse the dust particles inside these regions, see][]{Dullemond2018, Rosotti2020}. Most of these observations suggest that the typical $\alpha$ in protoplanetary disks is rather low \citep[see][for a review]{Rosotti2023}.

If disk viscosity is low, the question remains what is the main mechanism that drives the angular momentum transport and the evolution of protoplanetary disks. Knowing how the gas evolves in disks is fundamental to understand the dynamical behaviour of the particles. In fact, the gas disk mass and its distributions are poorly constrained from observations, but are crucial for determining the coupling of dust particles onto the gaseous disk.

\section{Disk Observations}
\subsection{Large-scale Substructures in Protoplanetary Disks}
\label{sect:substructures}
\begin{figure}
    \centering
    \includegraphics[width=\textwidth]{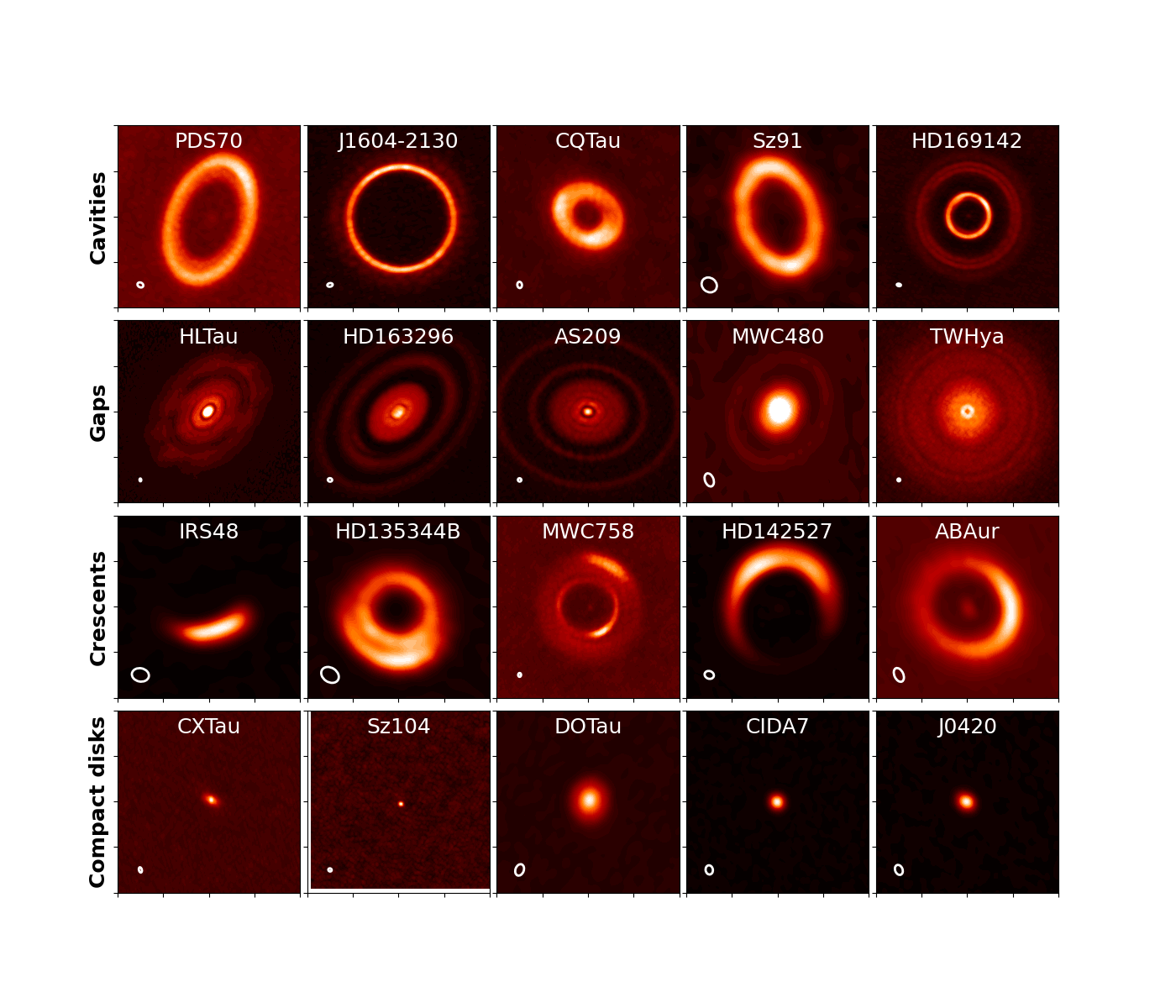}
    \caption{ALMA continuum gallery of examples of dust structures in protoplanetary disks, categorized in disks with inner cavities, gaps, crescents and compact disks. The beam size is indicated in the lower left corner. Images are 2"$\times$2" in size, except for HD142527 and AB Aur, which have been zoomed out to 4"$\times$4" as their angular size is larger. Individual disks: PDS70 \citep{Benisty2021}, J1604-2130 \citep{Stadler2023}, CQ Tau \citep{Ubeira2019}, Sz91 \citep{Mauco2019}, HD169142 \citep{PerezS2019}, HL Tau \citep{HLTau2015}, HD163296 \citep{Andrews2018}, AS209 \citep{Andrews2018}, MWC480 \citep{Liu2019}, TW Hya \citep{Andrews2016}, IRS48 \citep{vanderMarel2016-isot}, HD142527 \citep{Casassus2015}, AB Aur \citep{vanderMarel2021asymm}, MWC758 \citep{Dong2018obs}, CX Tau \citep{Facchini2019}, Sz104 \citep{vanderMarel2022}, DO Tau \citep{Long2019}, CIDA 7 \citep{Kurtovic2021}, J0420 \citep{Kurtovic2021}.}
    \label{fig:gallery}
\end{figure}

The previous section has demonstrated that \hbindex{dust traps} are expected to be present in disks to halt \hbindex{radial drift} and enable \hbindex{grain growth}. Such dust traps would appear as substructures in millimeter dust continuum images. High-resolution observations have revealed numerous substructures in the (sub)millimeter and optical/infrared images of protoplanetary disks, in particular in the \hbindex{ALMA} continuum emission tracing thermal dust emission at wavelengths between 0.5 and 3 mm, where the 1.3 and 0.85 mm (230 and 345 GHz or ALMA Band 6 and 7) wavelengths are used most commonly \citep{Andrews2020}. Dust disks in nearby star forming regions at 100-200 pc range in size from less than 10 to more than 100 au and their integrated continuum flux ranges from $\sim$1-100 mJy at 1.3 mm, which can be spatially resolved by ALMA in minutes. Substructures are seen in the form of e.g. rings, gaps, cavities, crescents, shadows and spiral arms, in clear contrast with the previously common assumption of smooth protoplanetary disks. Whereas the first large scale substructures were already seen in the pre-ALMA years in \hbindex{transition disks} with large inner cavities \citep{Brown2008, Brown2009, Andrews2011}, the \hbindex{HL Tau} disk that was observed at $\sim$0.04" spatial resolution ($\sim$6 au at 140 pc) as part of the ALMA Long Baseline Science Verification campaign \citep{HLTau2015} revolutionized the field revealing multiple narrow rings and gaps. This was followed by numerous other high-resolution ALMA studies of disks, most noticeably the DSHARP ALMA Large Program, which targeted 20 of the brightest protoplanetary disks at 0.04" resolution, revealing rings, gaps, crescents (asymmetries) and spiral arms at scales from a few to a few tens of au \citep[][and Figure \ref{fig:gallery}]{Andrews2018}. The distinction between a \hbindex{cavity} and a gap (see Figure \ref{fig:gallery}) is mostly historical, as cavities could already be identified through a deficit of infrared emission in the Spectral Energy Distribution (SED) and through imaging with pre-ALMA submillimeter interferometry \citep{Espaillat2014} whereas gaps leave no signature in the SED and can only be revealed through high-resolution imaging. In the context of this chapter, the definition of a cavity is thus an inner dust gap that is $>$20 au in size with little or no millimeter-dust inside following conventions in disk literature, whereas a dust gap is located in between narrow dust rings throughout the disk. 

Substructures are thought to be connected with planet formation, but it is currently not yet clear whether they are the start or consequence of young protoplanets in the disk, or both. It is also unknown if all substructures, even for the same morphology, are caused by the same phenomenon, or whether this varies from disk to disk or perhaps per evolutionary stage. Rings, gaps and cavities are often associated with \hbindex{planet-disk interactions}, as (growing) protoplanets will carve a gap in the disk along their orbit \citep{Artymowicz1996}, where the depth and width of the gap depend on the planet mass and disk viscosity \citep[e.g.][]{Fung2016,Zhang2018}: these observable parameters are thus key in connecting them with planet-disk interaction models. Although a minimum planet mass is required to create a dust gap, also called the \hbindex{pebble isolation mass}, which is generally above a Neptune mass \citep{Lambrechts2014, Rosotti2016}, its value depends strongly on the assumed disk viscosity \citep{Ataiee2018}. On the other hand, if dust rings are the start of planet formation in dust traps (see previous section), other mechanisms not invoking planets may in fact be responsible for them, such as ice-lines, photoevaporation and numerous magneto-hydrodynamic (MHD) and other disk instabilities \citep[see previous section and ][for an overview]{Bae2022, Lesur2022}. In that case, local dust concentrations facilitate the formation of planetesimals and ultimately planets \citep{Youdin2005,Johansen2007} rather than being caused by them. Whereas ice-lines initially appeared to be promising explanations for dust rings \citep{Zhang2016}, later studies showed that this could at least not be a universal explanation as ring radii did not correlate with ice-line radii in larger samples of disks \citep{Huang2018,long2018,vanderMarel2019}. However,  temperature variations in the disk can be caused by different physical processes and instabilities \citep{Owen2020}. MHD instabilities remain challenging to test observationally, and may only be confirmed by statistical arguments. 

A proper test on the origin of gaps would be the direct detection of protoplanets in disks, but this remains observationally challenging with current instruments. \hbindex{Protoplanets} have been firmly detected in three disks so far through direct imaging: PDS~70 \citep{Keppler2018,Haffert2019}, AB~Aur \citep{Currie2022} and HD~169142 \citep{Hammond2023}, as well as upper limit constraints in various other systems \citep[e.g.][]{Asensio2021,vanderMarel2021asymm}. Models of dust evolution and planet-disk interaction have demonstrated that inferring the planet mass from the morphology of the substructures is complicated and degenerate, because the shape of the substructures depends on different disk parameters \citep{Bae2018}. Indirect signs of the presence of protoplanets in gaps have been found in numerous disks as well, for example through deep gas cavities \citep[e.g.][]{vanderMarel2016-isot}, \hbindex{non-Keplerian kinematics} in CO line cubes \citep[e.g.][]{Pinte2018}, pressure gradients \citep{Teague2018}, \hbindex{circumplanetary disks} \citep{Benisty2021, Bae2022cpd}, localized shock emission \citep{Booth2023} and spiral arms and shadows in scattered light images \citep[e.g.][for an overview]{Benisty2022}. Together with this indirect evidence, protoplanets seem to be a realistic explanation for at least the large scale gaps and cavities in disks, although other explanations (see previous section) cannot be excluded at this point for most systems. 
However, as dust concentrations are initially required to start the planet formation process, the chicken-and-egg problem of dust substructures remains.

Other than dust rings and gaps, other types of substructures have been observed regularly in disks as well even at lower resolution of 0.15-0.25", in particular crescents (asymmetries) and spiral arms, both in (sub)millimeter and scattered light images. \hbindex{Crescents} have been detected along millimeter dust rings in various disks, in a range of contrast values from 1.5-2 up to more than 100 \citep{Perez2014,Pinilla2014,vanderMarel2021asymm}. The first crescent disk, Oph IRS 48, provided robust observational evidence for the presence of dust trapping in disks (see previous section), by comparison of the distribution of the millimeter dust and gas and small dust grains \citep{vanderMarel2013}. Crescents are also found in secondary rings \citep{Cazzoletti2018} or as clumps in part of the ring \citep{Perez2018}. Crescents are thought to be azimuthal variants of the radial dust traps discussed above, possibly due to \hbindex{Rossby Wave Instability} leading to long-lived vortices \citep{BargeSommeria1995,KlahrHenning1997} or eccentric disks due to massive companions \citep{Ataiee2013,Ragusa2017}. Multi-wavelength observations tracing different grain sizes show different azimuthal widths \citep{Pinilla2015,vanderMarel2015,Casassus2015} and shifts in peak positions \citep{Cazzoletti2018}, consistent with modeling predictions \citep{Birnstiel2013,BaruteauZhu2016}. \citet{vanderMarel2021asymm} has proposed that mm-dust asymmetries only become visible at mm-wavelengths when dust rings are located at radii with low gas surface density (i.e. far out in the disk or in disks with low gas mass) to explain the diversity of crescents in disks but this remains to be proven with larger samples and better estimates of the gas surface densities in disks. Overall the sample of crescents in disks is rather small, with an occurrence less than 10\%, which makes it difficult to address their nature w.r.t. axisymmetric dust rings. 

Second, \hbindex{spiral arms} have been detected in both scattered light \citep[e.g.][]{Benisty2022} and in millimeter emission \citep{Perez2016,Huang2018}, and differences in appearance between the two wavelengths have been related to scale height \citep{Juhasz2015,Rosotti2020}, whereas the origin has been interpreted as either spiral density waves triggered by companions \citep{Dong2015spirals} or gravitational instabilities \citep{Kratter2016}, returning to the question in the beginning of this section: are substructures the result or start of planet formation? 

Substructures in molecular line emission, in particular CO isotopologues, have been detected as well with ALMA, in some cases connected with the observed dust substructures. Disks with large inner dust cavities $>$30 au often show gas cavities in $^{13}$CO and C$^{18}$O with radii that are 1-2 times smaller than their dust counter parts \citep[e.g.][]{vanderMarel2016-isot,Dong2017,Ubeira2019}, consistent with a planet gap model where dust is trapped at the outer edge of a gas gap \citep{Pinilla2012b,Facchini2017gaps}. On the other hand, the narrow gaps of $<$10 au width remain mostly undetected in CO line emission, due to a combination of spatial resolution and temperature changes inside a dust gap \citep{Isella2016,vanderMarel2019,Rab2020}. Radial substructure has been detected in images of other molecules in gapped disks, e.g. in the MAPS ALMA Large Program survey \citep{Oberg2021}, but without a clear correlation to the dust substructure \citep{Law2021,Jiang2022}, suggesting other physical mechanisms contribute to their gas abundance and excitation. Ice-lines and dust transport may play an important role in setting the molecular substructure, as molecules like N$_2$H$^+$ only become abundant when CO is frozen out, resulting in outer N$_2$H$^+$ rings \citep{Qi2013,Qi2019}, and ice-produced molecules like CH$_3$OH have been detected in disks with significant amount of warm dust at temperatures $>$100 K where these molecules sublimate \citep{Booth2021a,vanderMarel2021-irs48}. Furthermore, CO emission has revealed spiral structures in several disks after subtraction of the rotating disk component \citep{Christiaens2014,Tang2017,Teague2019twhya} similar to the scattered light images. Finally, CO line cubes show a range of non-Keplerian substructures which are thought to be linked to forming protoplanets as well, such as velocity kinks, warps, streamers and circumplanetary disks, which are not further discussed here, but are reviewed in detail by \citet{Pinte2022}.

\begin{figure}
    \centering
    \includegraphics[width=\textwidth]{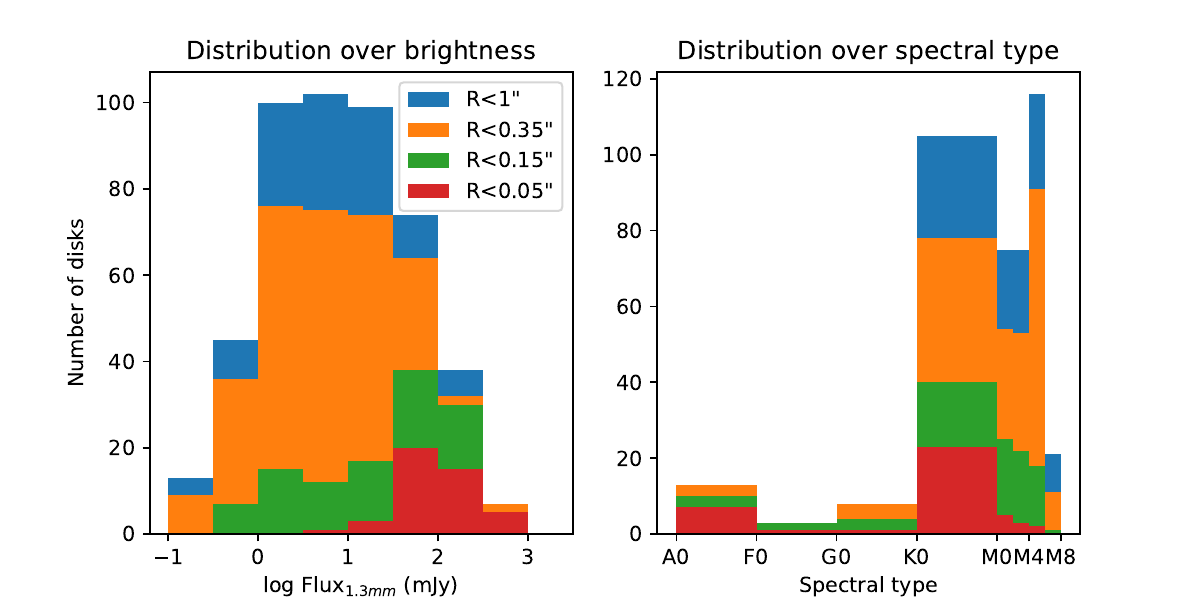}
    \caption{The distribution of the observed angular resolution of protoplanetary disks imaged with ALMA, using the data from \citet{vanderMarelMulders2021} from the Taurus, Lupus, Ophiuchus, Chamaeleon and Upper Sco regions as well as isolated Herbigs. In the left plot the distribution is shown w.r.t. the integrated flux, in the right plot w.r.t. spectral type. For clarity, the M-dwarfs are grouped per 2 subtypes. This plot demonstrates that high-resolution ALMA observations are relatively biased towards the brightest disks and the earliest spectral types, even though such disks are rare in the overall distribution.}
    \label{fig:resolution}
\end{figure}

\subsection{Observational Biases}
By today, it appears that almost any disk that is observed at sufficiently high angular resolution reveals some kind of substructure, but one has to be mindful of strong selection biases when drawing such conclusions. As the emission scales with the dust disk size \citep[e.g.][]{Tripathi2017}, the most extended disks are also the brightest, and therefore the most often selected for high-resolution observations, which naturally results in a higher detectability of substructures \citep{Bae2022}. Furthermore, the observed disk dust mass (see later discussion) scales with the stellar mass \citep[e.g.][]{Ansdell2016,Ansdell2017} and as the calculated dust mass is proportional to the integrated flux, the brightest disks are most commonly found around higher mass stars, introducing another bias. Whereas the prevalence of substructures on first sight appears to be inconsistent with the occurrence rates of giant exoplanets, biases have to be considered before drawing such conclusions. In particular, the shape of the initial mass function (IMF) which peaks around mid-M or subsolar mass stars, and its corresponding coverage in high-resolution ALMA studies, directly reveals the biases that still trouble observational disk studies (see Figure \ref{fig:resolution}). This figure also shows the bias towards brighter protoplanetary disks, making it clear that fainter, smaller disks are mostly unresolved in current ALMA observations at low spatial resolution, lacking any information on possible substructures. Disk populations and the attempts to correct for these biases are discussed at the end of this chapter.

\subsection{Evolutionary Stages}
Substructures in disks have been studied extensively in the protoplanetary disk or Class II stage at disks of 1-10 Myr as described in the previous section, but are also detected in the earlier, \hbindex{embedded disk} stage. This stage is essentially defined as the time before the envelope from which the star and disk have formed has fully dissipated, in the first Myr of the lifetime of the disk. Given the chicken-and-egg problem of dust substructures and planet formation described in the beginning of this section, it is a natural question whether substructures are already present in the embedded stage, and if so, if they have a similar nature and origin as their counterparts in the Class II stage. If this is the case and proven to be caused by growing protoplanets, this sets firm constraints on the time available to form planet cores that are massive enough to carve gaps: such cores would have to form in less than 1 Myr, which is challenging for current planet formation models. On the other hand, if substructures are not present or clearly of a different nature, this would imply that other mechanisms like MHD instabilities or gravitational instabilities create initial dust concentrations, which result in protoplanets that carve the gaps and dust rings that are detected in the protoplanetary disk stage.

ALMA observations have already revealed substructures in younger disks. In fact, the first highly resolved disk \hbindex{HL Tau} is a Class I disk, embedded in an envelope and only $\sim$1 Myr old, where narrow gaps and rings of only a few au wide were detected at 0.04" resolution \citep{HLTau2015}. Similar radial substructures have been seen in several other Class I disks observed at this resolution \citep{Sheehan2018,SeguraCox2020}, as well as inner cavities with radii $>$20 au \citep{Sheehan2017,Sheehan2020,Alves2020} and spiral arms \citep{Lee2020}. However, substructure is not seen as commonly and samples are not equivalent in completeness as in the Class II stage \citep{Tobin2020}. Recently, the eDISK ALMA Large Program targeting $\sim$20 Class 0 and I disks at 0.04" resolution revealed very few disks with substructure, which was interpreted as either rapid planet formation between the Class I and Class II stage, or obscuration of gaps by optical depth effects \citep{Ohashi2023}. The latter is supported by observations of \hbindex{edge-on disks} which directly reveal their vertical structure, showing that even though millimeter-dust is very settled in the Class II stage \citep{Villenave2020}, younger, embedded disks show more flared vertical structures at millimeter wavelengths \citep{Michel2022,Villenave2023}. Overall the early evolution of substructures in disks remains challenging to constrain, and samples of nearby disks that can be fully spatially resolved are limited to a few dozen. In contrast, the nearby low-mass star forming regions within 200 pc contain $\sim$1000 protoplanetary disks for which dust disk properties have been constrained with at least low-resolution imaging \citep{Manara2022}. 

During the protoplanetary disk evolution at 1-10 Myr, no clear evolutionary processes have been identified in substructures in disks, as substructures are seen both in the younger 1-3 Myr star forming regions as well as in the older 5-10 Myr regions. Biases and uncertainties in especially the older disk samples make evolutionary effects difficult to assess. Some studies indicate a possible evolutionary link between disks with gaps and disks with cavities, as a growing protoplanet carves a wider gap over time \citep{vanderMarel2018,Cieza2020}, but this cannot be proven with the available small samples and large variety of gap widths.

It is tempting to link the observed substructures in protoplanetary disks to the dusty rings observed in nearby \hbindex{debris disks} or `exoKuiper belts', due to their similarity in appearance in dust in the form of rings, gaps, cavities and asymmetries \citep[e.g. Fig. 2 in ][]{Marino2023}. However, the dust in debris disks is generated by planetesimal collisions and is no longer coagulating or affected by drag forces in the absence of gas \citep{Hughes2018}, making them fundamentally different from protoplanetary disks. Furthermore, they range in ages anywhere between 10 Myr and 10 Gyr, and their detection rate is heavily limited (and thus biased) by sensitivity, despite many of being much closer than 100 pc. In particular, they are primarily detected around early type stars also after correction for sensitivity \citep{Sibthorpe2018} and it is challenging to define complete samples as in protoplanetary disk surveys. 

When comparing debris disk properties with protoplanetary disks, the dust rings w.r.t. their radius tend to be wider, and a correlation between stellar luminosity and ring radius exists in debris disks, suggesting a link with their formation process, in contrast with dust rings in protoplanetary disks \citep{Marino2023}. One important argument for having a connection is the fact that debris disks require planetesimals to be located at tens of au, and that planetesimals at those radii can grow efficiently in dust traps \citep{Stammler2019,Pinilla2020, Miller2021}, whereas pebbles would otherwise drift inwards before growing to boulders and larger \citep[][]{Michel2021,Najita2022}. However, planetesimal belts need to be dynamically stirred to fragment into detectable debris levels, requiring giant planets inside their dust ring \citep{Wyatt2015}, and a statistical correlation between cold giant planets and debris disks has only been tentatively found so far \citep{Meshkat2017}.

\subsection{Inner Disk Morphology and Substructures}
\label{sect:innerdisks}
The \hbindex{inner disk} or inner 1 au of protoplanetary disks is particularly interesting in connection with exoplanets, as a large part of the known exoplanet populations (including the rocky planets in our Solar system) reside in this regime. The inner 1 au in protoplanetary disks cannot be spatially resolved with ALMA due to the distances of nearby young disks and the available spatial resolution, as well as the high continuum optical depth in the inner disk. The presence of inner dust disks has been inferred indirectly from near infrared excess in SEDs, in particular in disks with inner cavities \citep{Espaillat2014}. At these distances from the star, dust grains sublimate at $\sim$1500 K and the inner edge (sublimation radius) is directly irradiated by the star, resulting in a puffed up wall and resulting strong near infrared emission \citep{Dullemond2010}. Optical and infrared see-saw variability can be used to infer more information on azimuthal and/or temporal changes in the inner disk vertical structure due to time-dependent obscuration of the star \citep{Espaillat2011}. Inner dust disks have also been detected at millimeter wavelengths, indicating large grains may be present as well \citep{Francis2020}.

Another way of inferring more information on the structure of the inner dust disk is optical interferometry, e.g. with the \hbindex{VLTI} or CHARA arrays at near infrared wavelengths. Inner disk surveys with VLTI instruments such as PIONIER, GRAVITY and recently MATISSE have revealed the geometry of inner dust disks \citep{Menu2015,Lazareff2017,Perraut2019,Kluska2020,Varga2021,Bohn2022}. Whereas forward imaging is not possible in the same way as with submillimeter interferometry due to the limited uv sampling, radial profiles can be fit to the visibilities and deviations from axisymmetric emission (such as crescents, clumps or protoplanets) can be identified from the closure phase. Such studies are still limited to the brightest systems and samples are thus particularly biased. 

Alternatively, infrared molecular line emission originating from the inner region of the disk also provides valuable information about the substructure in the inner disk due to the sublimation of ices frozen out on dust grains that have been transported to the inner part of the disk. The \emph{James Webb Space Telescope (\hbindex{JWST})} is expected to revolutionize this particular field of disk studies, as its sensitivity and spectral resolution exceeds that of \emph{Spitzer}, which means that many additional molecular species will be detectable in larger disk samples. A tentative anti-correlation between infrared H$_2$O emission as detected with \emph{Spitzer} and disk dust size already hints at the importance of dust transport in setting the chemical composition in the inner part of the disk \citep{Banzatti2020,Kalyaan2021}. Several first result \emph{JWST} studies indeed show that infrared molecular detections can probe inner disk structures and reveal more about their disk evolution history \citep{Grant2023,Tabone2023,Banzatti2023}. A thorough understanding of dust evolution and transport in disks is crucial for a proper interpretation of these new data.


\section{Disk Dust Masses}
\label{sect:dustmass}
One of the most fundamental properties of the disk is its total mass, which sets the budget for the planets that can be formed in the disk and that governs the general dynamics in the star-disk system, including gravitational instability. The disk mass has historically been computed as the \hbindex{Minimum Mass Solar Nebula} of $\sim$60 $M_{\oplus}$ in solids, 
using the amount of material in the Solar System planets \citep{Hayashi1981}, but can now be derived from protoplanetary disk observations directly, showing a large range in values. Whereas the gas mass is thought to be the dominant contributor at 99\% of the disk mass, the \hbindex{dust mass} is easier to constrain observationally, and is therefore often used as proxy for the disk mass, assuming a certain gas-to-dust ratio (see end of this section). The dust mass of the disk is generally computed using the integrated millimeter flux, under several assumptions, as follows \citep{Hildebrand1983}.

The integrated flux $F_{\nu}$ can be described as the specific intensity $I_{\nu}$ integrated over the solid angle d$\Omega$ of the disk:
\begin{equation}
\label{eqn:fnu}
F_{\nu} = \int{I_{\nu}d\Omega}
\end{equation}
where $I_{\nu}$ is computed from the Planck curve at an assumed temperature $T$:
\begin{equation}
\label{eqn:bnu}
    I_{\nu}=B_{\nu}(T)(1-e^{-\tau_{\nu}})
\end{equation}
where the optical depth $\tau_{\nu}$ is proportional to the surface density $\Sigma_{\rm dust}$ and the \hbindex{dust opacity} $\kappa_{\nu}$ and scaled with the inclination i:
\begin{equation}
    \tau_{\nu}=\frac{\kappa_{\nu}\Sigma_{\rm dust}}{\cos i}
\end{equation}
Since the dust mass $M_{\rm dust}$ is the integrated surface density
\begin{equation}
    M_{\rm dust} = \int_0^R \Sigma_{\rm dust}(r)2\pi rdr
\end{equation}
with $R$ the size of the disk, and using 
\begin{equation}
    d\Omega = \frac{\pi R^2}{d^2}
\end{equation} 
with distance $d$ this can be rewritten as
\begin{equation}
    M_{\rm dust} = \frac{d^2F_{\nu}}{\kappa_{\nu}B_{\nu}(T)}
\end{equation}
under the assumption $\tau_{\nu}\ll 1$ so that 1-e$^{-\tau_{\nu}}\approx \tau_{\nu}$, implying that the emission is optically thin and thus represents the entire column. The dust opacity $\kappa_{\nu}$ depends on the grain size distribution and properties \citep{Draine2006} but is generally assumed as $\kappa_{\nu}$=10 cm$^2$ g$^{-1}$ at 1000 GHz \citep{Beckwith1991} and scales as $\kappa_{\nu}\sim\nu^{\beta_{mm}}$ with dust opacity index $\beta_{mm}$=1, hence $\kappa_{\nu}$=2.3 cm$^2$ g$^{-1}$ at 230 GHz. For the temperature, T=20 K is usually assumed for all disks as a scaling with the stellar luminosity has been shown to make marginal differences \citep{Andrews2013}.

This dust mass calculation is generally used in large disk surveys due to the simple relation with the observed millimeter flux and distance, and can be used to compare (evolutionary) behaviour across samples. Full radiative transfer models that are finetuned to specific disk structures lead to small differences \citep{AndrewsWilliams2007tau}, which can be up to a factor of 2-3 in the presence of large cavities \citep{vanderMarel2018}. 

\subsection{Dust Opacity and other Uncertainties}
\label{sect:opacity}
The computed dust mass thus relies heavily on several assumptions: the dust temperature T, the assumed grain properties that set $\kappa_{\nu}$ and the optically thin assumption. They will be discussed one by one.

The \hbindex{dust temperature} is generally assumed to be 20 K, under the assumption that the bulk of the dust mass is within that temperature regime. The disk midplane temperature can be approximated by Eqn. \ref{eq:temp}.  
One can show that the temperature is $\sim$15-20 K outside 20 au for typical stellar luminosities between 0.1 and 1 $L_{\odot}$, so if the dust in the disk is typically distributed out to 100 au this is a reasonable assumption for the dust mass calculation. Furthermore, in the Rayleigh-Jeans approximation which is typically valid in the millimeter regime ($h\nu\ll k_BT$) the Planck distribution $B(T)$ can be approximated by:
\begin{equation}
\label{eqn:rj}
    B_{\nu}(T(r))=\frac{2\nu k_BT(r)}{c^2}
\end{equation}
so the dust mass scales inversely linear with the assumed temperature. It is important to note that if the disk dust size is much smaller than 20 au (which may well be the case for the observed fainter disks, consistent with radial drift), the temperature and thus the dust mass is  overestimated by a factor of a few under the assumption of optically thin emission, as illustrated in Figure \ref{fig:luminosity}). However, if the dust is optically thick (see below) this effect may be compensated.

\begin{figure}
    \centering
    \includegraphics[width=\textwidth]{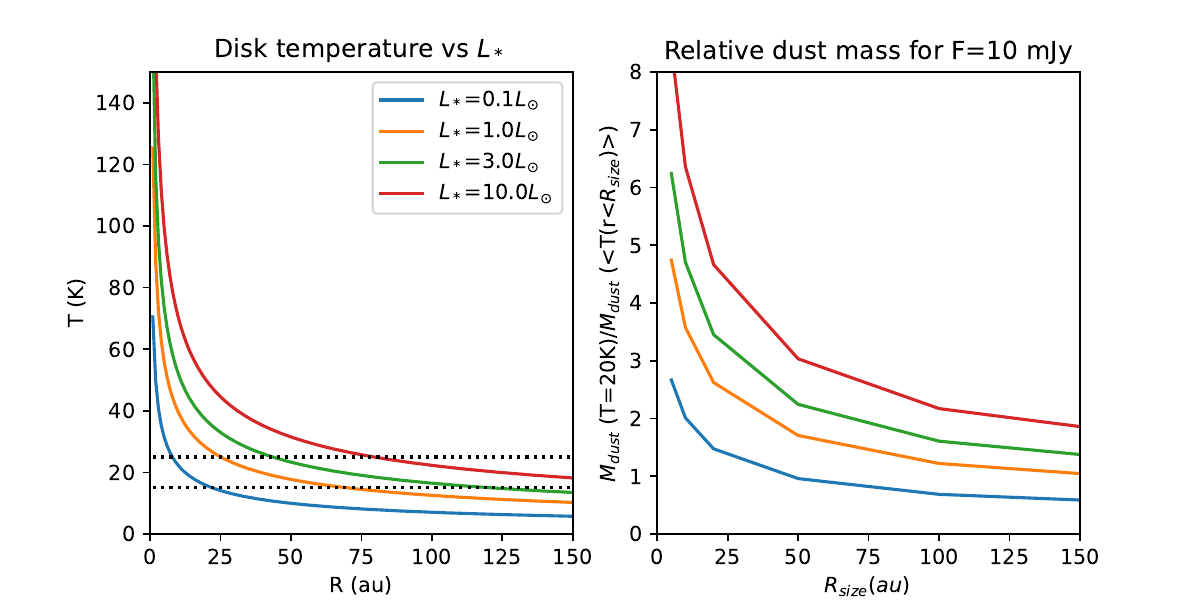}
    \caption{The effect of the chosen disk temperature on the computed dust mass. The left panel shows the temperature curve from Eqn. \ref{eq:temp} for different stellar luminosities. The 15-25 K regime is indicated by dotted lines, showing that the bulk of the outer disk is located around 20 K if the disk is larger than 20 au. The plot on the right shows how much the classical dust mass calculation with T=20 K overestimates the actual dust mass assuming the average temperature in the disk for different disk sizes for a flux of 10 mJy at 230 GHz. Especially for disks $<$20 au in size the classical dust mass is a strong overestimate, which may be compensated by the presence of optically thick emission.}
    \label{fig:luminosity}
\end{figure}

The \hbindex{dust opacity} is set by the grain size properties, e.g. the \hbindex{maximum grain size} $a_{\rm max}$, the power $q$ in the size distribution $n(a)\propto a^{-q}$ and the dust composition and porosity \citep{Birnstiel2018}. Several dust opacity tables can be found online, or generated by the user for given properties (see \url{https://github.com/birnstiel/dsharp_opac}). The main issue here is that (much) larger grain sizes than the observing wavelength will not significantly contribute to the emission, as the primary contributor at observing wavelength $\lambda_{\rm obs}$ are grains with size $a=\lambda{\rm obs}/2\pi$ \citep{Draine2006}. This means that if a large amount of grains have grown to boulders and planetesimals (see previous section), their total dust mass is larger than the observable dust mass at millimeter wavelengths. The significance of this effect can be as large as two orders of magnitude \citep[][]{Pinilla2020}.

The third uncertainty is the assumption of optically thin emission. If the emission is optically thick, $\tau_{\nu}>$1, and Eq. \ref{eqn:bnu} simplifies to $I_{\nu}=B_{\nu}$ and the flux no longer scales with $\Sigma_{\rm dust}$, but with $T$. In reality, both density and temperature  contribute to the emission, but it implies that the dust mass is underestimated using the observed flux alone. As the \hbindex{optical depth} decreases with wavelength, fluxes measured at 3 mm and longer can be more realistically assumed to be optically thin. Multi-wavelength studies including integrated fluxes at longer wavelengths have demonstrated that the amount of optically thick emission is limited in disks and thus does not underestimate the dust mass by more than a factor of a few \citep{Tychoniec2020,Tazzari2021}, as most of the optically thick emission is located in the center of the disk when spatially resolved  \citep[e.g.][]{Carrasco2019,Macias2021,Guidi2022}. 

Another consequence of optically thick emission is that the dust opacity index $\beta_{mm}$, which is generally used to constrain \hbindex{grain growth} (see next section) can no longer be derived from the observed spectral index $\alpha_{mm}$. In the Rayleigh-Jeans (Eqn. \ref{eqn:rj}) and optically thin regime, the flux density from Eqn. \ref{eqn:fnu} scales as
\begin{equation}
\label{eqn:spindex}
F_{\nu}\propto B_{\nu}\tau_{\nu}\propto\nu^2\kappa_{\nu}\propto\nu^{2+\beta_{mm}}
\end{equation}
so the observationally constrained spectral index $\alpha_{mm}$ from $F_{\nu}\propto\nu^{\alpha_{mm}}$ equals 2+$\beta_{mm}$. However, if the emission is optically thick this relation is no longer valid and the grain size can no longer be constrained from the observed spectral index.

\subsection{Solid Mass Budget} 
\label{sect:mass_budget}
Several disk studies have estimated the solid \hbindex{mass budget} available for planet formation during the protoplanetary disk phase using dust continuum observations of large disk samples, using the \hbindex{dust mass} derivation above. These dust masses are found to be significantly lower than the solid mass in observed exoplanet cores, suggesting that either planet formation has already finished by the protoplanetary disk phase or that the observed dust masses are severely underestimated \citep{Manara2018,Tychoniec2020}. A more recent analysis where  exoplanet selection and detection biases are taken into account finds that the disk dust masses are in fact comparable to the exoplanet core masses, which would suggest a 100\% efficient conversion of dust mass into exoplanet cores, and thus inconsistent with most planet formation models \citep{Mulders2021a}. Although the previous section demonstrates that the dust mass in disks could be underestimated, it is unclear if this is sufficient to explain this discrepancy with exoplanet masses. The alternative explanation is that protoplanetary disks are actually remnants of planet formation, and just reveal the dust that is left. This dust may still grow into planetesimals and form Kuiper belts or asteroid belts. In that case, dust masses of embedded disks rather than protoplanetary disks should be used in planet formation models as initial conditions, which are at least 1 order of magnitude more massive. 

\subsection{Dust Mass Evolution and Trends}
Disk dust masses do not only provide initial conditions, but can also be studied in evolutionary context to understand the main processes that dissipate them \citep{Manara2022}. Protoplanetary disk surveys reveal that the general dust mass distribution decreases over time from the Class 0 to the Class III and debris disk stage over several orders of magnitude \citep[][and Figure \ref{fig:evolpathways}]{Cieza2019,Williams2019,Michel2021}. As the dust mass is correlated  with the dust disk size \citep{Tripathi2017}, it is not surprising that the observed disk dust size decreases with age as well during the Class II phase \citep{Hendler2020,Tazzari2021}. Some correlation exists between the dust mass and the disk accretion rate \citep{Manara2016}, but the scatter in this correlation is large, casting doubts on their physical relation \citep{Manara2020}. 

\begin{figure}[!ht]
    \centering
    \includegraphics[width=0.6\textwidth]{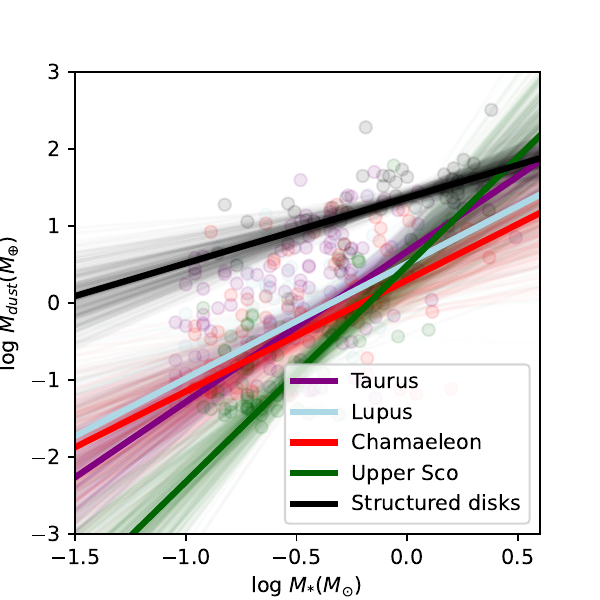}
    \caption{Observed correlation between dust mass and stellar mass, using data from ALMA surveys of nearby star forming regions between 1 and 10 Myr. Note that Ophiuchus disks \citep{Cieza2019} are not included as the known stellar masses are incomplete. The different regions are color-coded according to the legend. Transparent circles represent the data points, the solid lines represent the best-fit linear correlations using the \texttt{linmix} python package to include upper limits, with the grey lines representing the spread in the fit. The structured disks containing dust traps (disks with cavities and gaps) are grouped separately from the regions. All data points \citep{vanderMarelMulders2021} are corrected for their Gaia DR2 distance. The plot shows that the correlation between dust mass and stellar mass steepens with time \citep{Ansdell2017} but structured disks with dust traps follow a flatter relation \citep{Pinilla2020}.}
    \label{fig:stmass}
\end{figure}

Furthermore, the dust mass correlates with the stellar mass, albeit with large scatter, although the slope of this correlation steepens over time, indicating that the dust mass decreases more rapidly for lower mass stars \citep[][and Figure \ref{fig:stmass}]{Ansdell2017}.  When disks with dust rings and gaps are considered separately, their correlation with stellar mass is much flatter than the majority of disks \citep{Pinilla2018tds}. The main correlation implies that higher mass stars generally have higher mass disks, but considering the scatter in this relation, it should be noted that a small fraction of disks around low-mass stars (M-dwarfs) do have higher mass disks, perhaps even massive enough to form giant planets or to contain remnants of giant planet formation \citep{Curone2022}.

Whereas there is still active debate on the general mechanism of gas dissipation in the disk (viscously driven or driven by a \hbindex{magnetohydrodynamical wind}, or both, see \citet{Manara2022}), in recent years it has been proposed that the dust evolution is decoupled from the gas disk evolution \citep{Sellek2020}. Disk models including radial drift (see first section) can reproduce the observed trends in dust mass for the majority of disks, including the stellar mass dependence \citep{Pinilla2020,Zormpas2022,Appelgren2023}. This implies that the observed dust mass is not representative for the disk mass, and \hbindex{gas-to-dust ratio}s may be much larger than the ISM ratio of 100 for such disks (see later discussion), if the \hbindex{radial drift} is well on its way by the time the dust mass is observed. On the other hand, the amount of millimeter-dust in pressure bumps does not decrease by radial drift and the dust mass would remain high, but this would be only a small fraction of the full disk population considering the general trends. This would explain the massive outliers with dust traps seen in some older star forming regions \citep{Ansdell2020}. A scenario of separate evolutionary pathways for disks with and without \hbindex{dust traps} has been proposed \citep[][and Figure \ref{fig:evolpathways}]{vanderMarelMulders2021}, but remains to be confirmed by more uniformly observed disk data.

\begin{figure}[!ht]
    \centering
    \includegraphics[width=\textwidth]{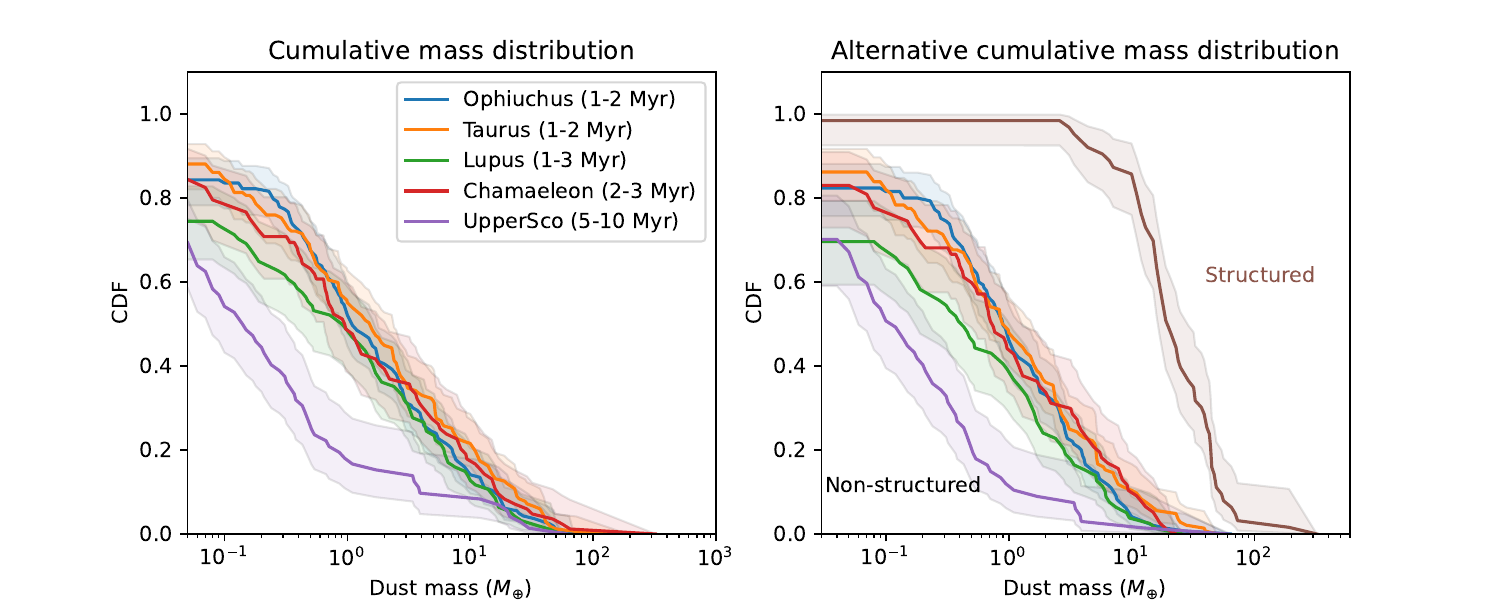}
    \caption{Cumulative mass distribution function of dust masses of Class II disks, using data from ALMA surveys of nearby star forming regions between 1 and 10 Myr, computed with the \texttt{lifelines} python package to take into account upper limits. The left plot shows the result of the full disk populations in each region, similar to \citet{Cieza2019}, the right plot considers the disks with structure (disks with cavities and gaps) and without structure separately, following \citet{vanderMarelMulders2021}. These plots demonstrate that the dust mass of the majority of disks decreases with age as the result of radial drift, whereas the dust mass of structured disks stays on the high end of the distribution.}
    \label{fig:evolpathways}
\end{figure}

\subsection{Gas-to-Dust Ratio}
\label{sect:gdr}
The observed dust mass is often used to compute a gas mass by multiplying with the ISM \hbindex{gas-to-dust ratio} of 100, to infer the total disk mass. Alternative methods using molecular line observations such as HD and CO isotopologues to infer the gas-to-dust ratio suggest that the ratio may be at least an order of magnitude lower, either due to increased gas dissipation, inaccuracies in the conversion from CO to H$_2$, or both. A full discussion of these uncertainties is beyond the scope of this chapter, but a review on the relevant aspects and consequences for disk gas masses has been written by \citet{Miotello2022}. Furthermore, the uncertainties in the dust mass calculation and the dust evolution processes cast further doubt on the validity of using a universal gas-to-dust ratio to compute disk gas masses.


\section{Dust Growth} \label{sect:dust_growth}
\subsection{Spectral Index and Grain Sizes}
\label{sect:spindex}
Submillimetre and microwave emission from the diffuse interstellar dust have shown that the typical values for the \hbindex{dust opacity index} $\beta_{\rm{mm}}$ (see previous section) in this media is 1.7-2.0 \citep{Finkbeiner1999}. Understanding how the opacity index varies in protoplanetary disks has been a driver for multi-wavelength observations in the last decade. If the dust in protoplanetary disks has a similar size distribution as in the interstellar medium, the opacity index should be similar. However, it has been found that the opacity index is (much) lower than 1.7, suggesting that dust grains are growing in disks \citep{Draine2006}.

The usual method to obtain the opacity index of protoplanetary disks is by assuming that the (sub-) millimeter emission is optically thin and in the Rayleigh-Jeans limit is directly proportional to the observed millimeter \hbindex{spectral index} $\alpha_{\rm{mm}}$, such that $\beta_{\rm{mm}}=\alpha_{\rm{mm}}-2$ (Eq. \ref{eqn:spindex}). By measuring the total millimeter flux of protoplanetary disks, the spatially integrated $\beta_{\rm{mm}}$ is less than 1.7 for disks in different star forming regions, for a large range of disk luminosities and types of stars. This indicates that large millimeter-sized particles are present in protoplanetary disks with different properties \citep[e.g.,][]{Beckwith1991,Testi2001, Ricci2010, Guilloteau2011, Tazzari2021b}. High angular resolution observations that are sufficient to resolve spatial variations of the spectral index, have shown that it increases in the outer disk (as expected from dust evolution models, where the larger grains are located in the inner parts \citep[e.g.,][]{Perez2015, Tazzari2021b}, and the spectral index seems to decrease inside disk substructures, suggesting that these regions are preferential for grain growth \citep[e.g.,][]{Casassus2015, vanderMarel2015, Carrasco2019, Guidi2022}. Current interpretations of the spectral index are limited because the emission in submillimeter images is not fully optically thin and because the effect of dust self-scattering is neglected.

\subsection{(Sub-)millimeter-wave Polarization}
\hbindex{Polarization} at sub-millimeter wavelengths has been detected in young disks (Class 0-I) as well as in more evolved disks (Class II) \citep{Rao2014, cox2015, kataoka2017}.
This polarization is usually interpreted by grain alignment perpendicular to the magnetic field, radiation flux, or by dust self-scattering \citep[][for a review]{anderson2015}. The polarization patterns by dust self-scattering can also provide insights about the grain size in protoplanetary disks.

This can be understood as follows: when dust grains in protoplanetary disks emit thermal radiation, this emission can be scattered by other dust grains, hence called 'self-scattering'.  Dust grains that are comparable to or larger than the observing wavelength, $a>\lambda/ 2\pi$, are expected to have a large albedo, so that their scattering efficiency is high enough to produce scattered emission. An efficient polarization is reached for $a\sim \lambda/ 2\pi$ because for grains much larger than the observing wavelength the scattering is strongly forward peaked, and no polarization is expected. Polarization observations at sub-millimeter wavelengths of a handful of disks suggest that the maximum grain size is around 100-150\,$\mu$m \citep[e.g.,][]{kataoka2016, kataoka2017}, but when multi-wavelength polarization maps are analysed,  it is difficult to find a single mechanism that could explain the detected polarization in protoplanetary disks \citep{Tang2023} and provide a maximum grain size. Additionally, the interpretation of millimetre polarization observations can become more complex in the presence of pressure bumps \citep[][]{pohl2016}.

\section{Exoplanet Populations and Trends w.r.t. Disks}
\label{sect:populations}
Exoplanets are generally grouped by their mass and location in different populations, each with their own trends on their \hbindex{occurrence rate} w.r.t. stellar mass, metallicity and orbital radius, for example super-Earths and sub-Neptunes, gas giants, wide-orbit (super-)Jupiter planets and rocky planets. As the exoplanet population is so diverse, the disk population should reflect this diversity as well if planet formation dominates the disk processes and their resulting appearance, at least after corrections for detection and selection biases  \citep{Drazkowska2022}. 

\subsection{Gapped Disks and Giant Planets}

One of the most obvious connections are the disks with large inner gaps w.r.t. wide orbit giant planets from high contrast imaging, as deep gaps require massive protoplanets or at least massive planet cores at large orbital radii at tens of au to carve them. Initial comparisons between the fraction of transition disks with inner cavities $>$20 au and the fraction of wide orbit giant planets at 10-100 au indicated strong discrepancies as the transition disk fraction of 11$\pm$3\% is significantly higher than the limits on wide orbit super-Jupiter occurrence of $<$4\% for FGK stars \citep[e.g.][]{vanderMarel2018}. Also for disks with narrow gaps at tens of au for which a sub-Jupiter mass is sufficient \citep{Lambrechts2019}, the giant exoplanet population appears to be inconsistent as the occurrence of giant exoplanets peaks at 2-6 au and drops to a few \% at larger radii \citep[e.g.][]{Fulton2021}. The obvious solution to this problem is inward \hbindex{migration} of these planets after they have created the gaps observed in disks to the location where they are observed today in the exoplanet population \citep{Lodato2019}. Even then, high-resolution disk observations appear to suggest that gapped disks are much more numerous than unstructured disks, inconsistent with the typical occurrences of giant exoplanets of $\sim$20-25\% \citep{Fernandes2019,Fulton2021}. 

The observational bias in protoplanetary disks studies (see Figure \ref{fig:resolution}) may play an important role here though, as discussed previously, as primarily brighter disks around higher mass stars have been imaged at high angular resolution \citep{Bae2022}. For example, the DSHARP survey discovered substructure in all disks imaged at 0.04" resolution \citep{Andrews2018}, but pre-selected disks that had been previously spatially resolved by ALMA and were thus by definition extended and bright. The unbiased Taurus disk survey by \citet{Long2019} suggested that around half of the disks around single stars contain substructure, but this survey only considered stars with spectral type $<$M2 (stellar masses $>$0.5 M$_{\odot}$).

\begin{figure}
    \centering
    \includegraphics[width=0.45\textwidth]{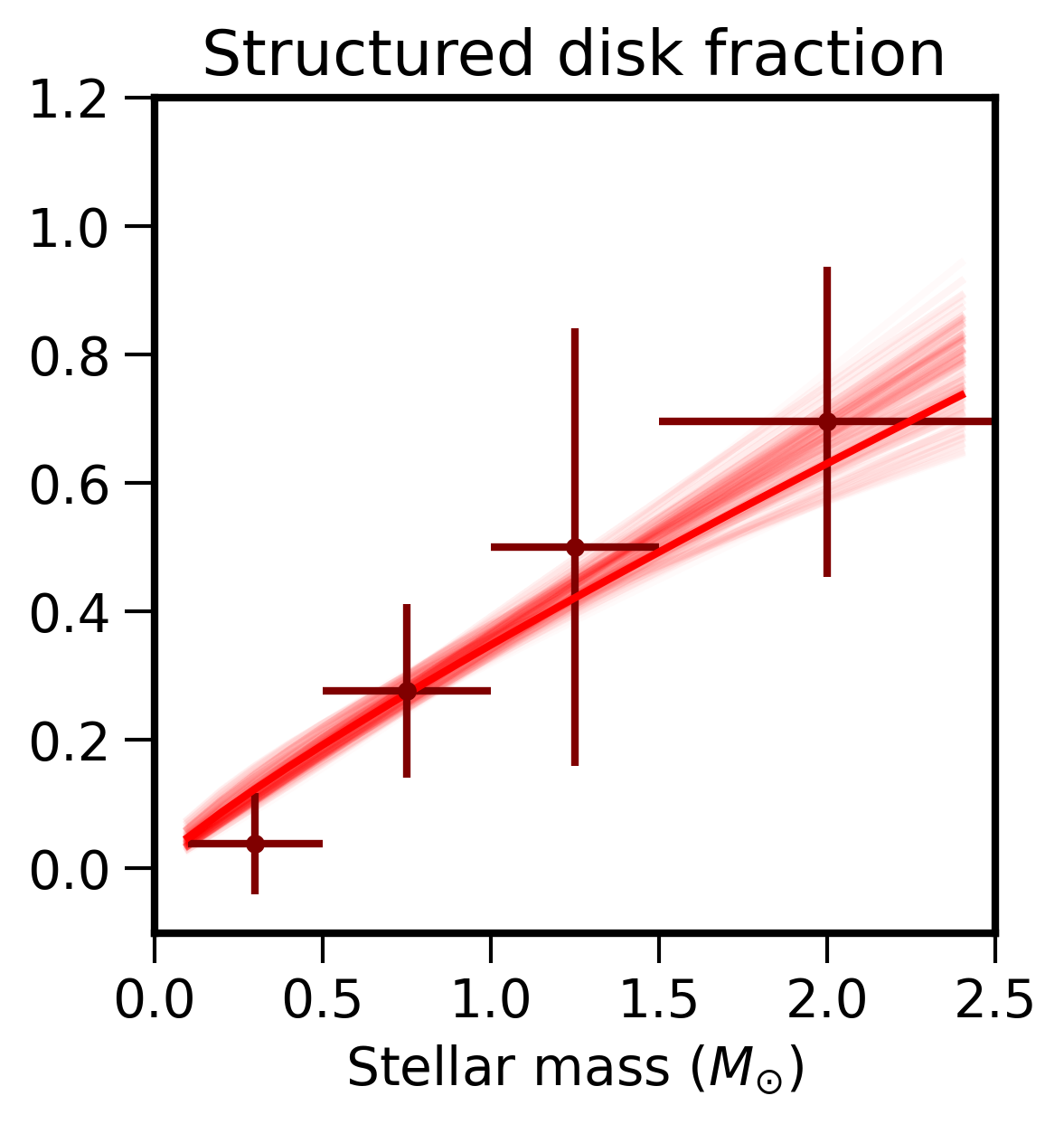}
    \includegraphics[width=0.45\textwidth]{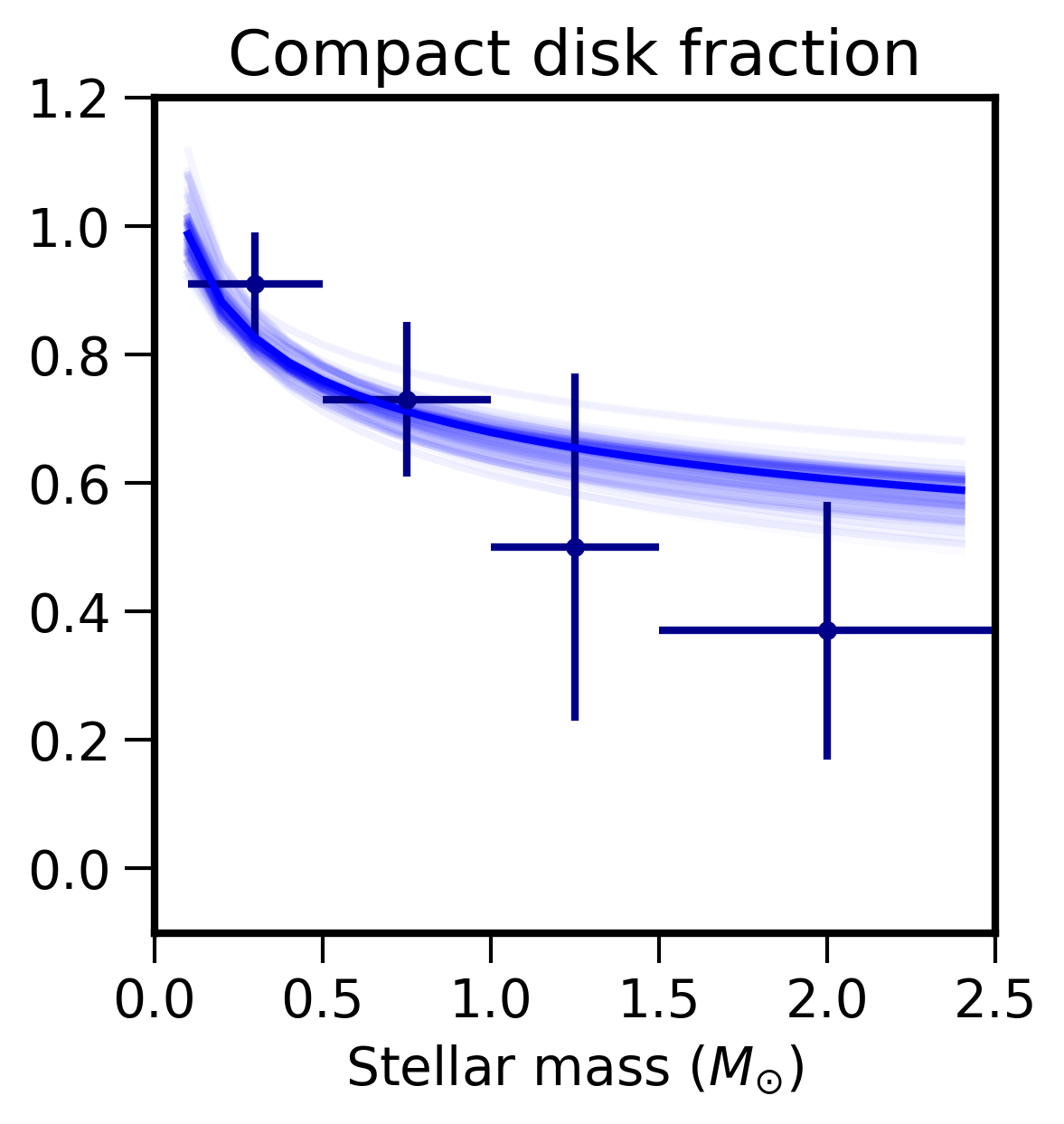}   
    \caption{The stellar mass dependence of disks with large-scale substructure and that of compact disks, based on the data from \citet{vanderMarelMulders2021}. These plots follow their Figure 7 for the combined structured disks (including transition disks, gapped disks and extended disks $>$40 au) and the compact disks (defined as disks $<$40 au in radius without observed substructure), not yet computed there. The plot shows a comparison between the computed disk fractions at pre-defined stellar mass bins (points with error bars) and the disk fraction estimated from a power-law fit (solid lines). The transparent lines indicate the spread in MCMC samples from the fit. The disk sample contains more than 500 disks from nearby star forming regions observed at a range of angular resolutions. The plot demonstrates that gapped disks are more common around high-mass stars whereas compact disks are most common around M-dwarfs.}
    \label{fig:stellarmassdep}
\end{figure}

Estimates of the fraction of disks that contain large-scale gaps in a very large disk survey \citep{vanderMarelMulders2021} range from 10-20\% when considering stellar masses from 0.1 to 3 M$_{\odot}$, much closer to the giant exoplanet occurrence rate, with the caution that not all disks have been sufficiently spatially resolved to rule out gaps. More specifically, they revealed a stellar mass dependence of the gapped disks for stars between 0.1 and 2.5 M$_{\odot}$, ranging from a few \% for M-dwarfs up to more than 60\% for stars $>$1.5 M$_{\odot}$, very similar to the stellar mass dependence of giant exoplanets (Figure \ref{fig:stellarmassdep}). A stellar mass dependence was also found for a sample of transition disks with large inner cavities \citep{vanderMarel2023} and a Herbig survey shows a fraction of more than 50\% of gapped disks, much higher than that of T Tauri stars \citep{Stapper2021}. 

The main shortcoming of these surveys is the lack of knowledge on substructures on smaller scales, in particular in the fainter disks. Only uniform high-resolution observations of full disk samples can fully confirm the validity of the stellar mass dependence. New imaging techniques have revealed subtle substructures in the inner part of some smooth, compact disks \citep{Jennings2022}. Also, at least one bright protoplanetary disk showed no evidence for substructure at 0.05" despite being radially extended \citep{Ribas2023}. Fully complete surveys at high angular resolution should provide better clues on the disk demographics and the significance of these trends.

\subsection{Compact Disks and Super-Earths}
A stellar mass dependence of gapped disks implies an opposite trend for drift-dominated disks: disks around low-mass stars are more likely to contain no large-scale dust traps and are thus more often very compact ($<$20 au) in size (Figure \ref{fig:stellarmassdep}). Furthermore, the drift efficiency has been shown to be more efficient around low-mass stars, as discussed in the first section of this chapter. Such disks often remain unresolved in low-resolution ALMA disk surveys, with upper limits on the dust size radius of $<$15 au up to at most $\sim$40 au, which makes it challenging to confirm their nature as smooth, compact disks without substructure. Only a handful of \hbindex{compact disks} has been radially resolved at a few au resolution with sizes of 3-10 au radius \citep{Facchini2019,Kurtovic2021,vanderMarel2022}. Their existence may be most evident from inner disk molecular content: if pebbles drift inward the H$_2$O snowline which is typically inside 1 au, their icy layers evaporate and increase the gaseous H$_2$O content, which can be measured with infrared facilities like \emph{Spitzer} and \emph{JWST} \citep{Kalyaan2021, Kalyaan2023}. An increased H$_2$O content has indeed been measured for compact dust disks \citep{Banzatti2020,Banzatti2023}, confirming their pebble drift nature. 

The population of drift-dominated disks could be natural sites for close-in super-Earth formation, due to the increased pebble rates in the inner part of the disk as a result of the radial drift \citep{Lambrechts2019,Mulders2021b}. Close-in super-Earths ('\hbindex{Kepler planets}') are particularly common in M-dwarfs \citep{Mulders2018}, following thus the trend seen in disks. Such planets can only be formed efficiently when the drift is not halted by the presence of a cold giant planet further out. This may contradict with the tentative correlation between close-in super-Earths and cold giants \citep{ZhuWu2018,Bryan2019,Fulton2021}, but this trend remains to be confirmed for low-mass stars.  


\subsection{Stellar Metallicity}
A second important trend in exoplanet populations is the correlation between the occurrence rate of giant planets with \hbindex{metallicity} \citep{Fischer2005}, which disappears for super-Earths and sub-Neptunes \citep{Buchhave2014,Kutra2021}. Whether such a trend with metallicity can be related to protoplanetary disks remains a topic of debate \citep{Drazkowska2022}.

Elemental abundances of the hosting stars are expected to be sensitive to different disk evolution processes, in particular efficient radial drift. Dust drift can enrich the stellar abundance of refractory elements at early times \citep{Huhn2023}. However, if trapping of dust particles occurs blocking the inward drifting particles  (for example by a giant planet), the stellar refractory elements should  decrease. The current correlation between more metal rich stars likely hosting more massive planets indicates that either the effect of the formation of the giant planet planets in their parental disk on the stellar photosphere was erased with the stellar evolution, possibly by the mixing of their convective envelope; or the observed correlation with the present-time values was stronger at earlier times. 

Stars more massive than the Sun,  have a radiative envelope rather than a convective envelope, which leads to slower mixing. Hence, for these stars ($\gtrsim $1-1.4\,$M_\odot$), the stellar photospheres can be dominated by the recent accretion of material and thus these stars are good candidates to test enrichment of refractory elements by dust drift.  \cite{Kama2015} demonstrated that there is a correlation between the presence of large disk cavities (that are likely caused by dust trapping) and the depletion of refractory elements of a set of Herbig Ae/Be stars. This conclusion was confirmed in \cite{Kama2019} by measuring the refractory sulfur in the photospheres of a sample of young stars with a mass of $>1.4\,M_\odot$, funding that Sulfur abundance is lower when trapping happened, presumably by a giant planet. For these stars, the inner disk abundance should be reflected in the stellar photosphere, which is supported by the results from \cite{Banzatti2018} who finds that the NIR excess of protoplanetary disks strongly relates to the Fe/H ratios in stellar atmospheres. 

The expected correlations between stellar metallicity and the type of planets forming around should depend then on the type of star (stars with mass lower than $\sim1\,M_\odot$ have a convective envelope and hence faster mixing), but also a dependency with the drift efficiency is expected, which is connected with the open questions of when giant planets form, the disk viscosity, and the micro-physical properties of dust particles, such as their fragmentation velocity.

\section{Outlook}
The revolution of observational work in protoplanetary disk studies has resulted in numerous new insights in disk processes and planet formation. State-of-the art models can now be compared much more directly to observations, revealing clues to the processes happening in these disks. As demonstrated in this chapter, the field of protoplanetary disks still suffers from several observational biases, and key for future development will be more uniform studies of larger samples across the IMF for proper comparisons with exoplanet properties. Higher sensitivity, as provided by ALMA in the Wideband Sensitivity Upgrade in the 2030s, will allow more rapid imaging of even the faintest disks in the nearby Universe. The \emph{James Webb Space Telescope} has only just started to deliver observations tracing the chemical composition in the warm regions of the disks, which may be highly important in constraining the dust structure in the terrestrial regime of the disk. Ground-based optical/infrared facilities are expecting major upgrades increasing sensitivity and improvements in the adaptive optics system and the Extremely Large Telescope (ELT) will provide another jump in sensitivity and resolution in infrared imaging of disks. At longer wavelengths, the next generation Very Large Array (ngVLA) and the Square Kilometer Array (SKA) are expected to trace the distribution of larger, centimeter-sized pebbles to shed further light on the dust growth and planet formation processes in protoplanetary disks. 



\section{Cross-References}
\begin{itemize}
\item{Imaging with Adaptive Optics and Coronographs for Exoplanet Research} 
\item{Planet Populations as a Function of Stellar Properties}
\item{A Brief Overview of Planet Formation}
\item{Chemistry During the Gas-Rich Stage of Planet Formation}
\item{Instabilities and Flow Structures in Protoplanetary Disks: Setting the Stage for Planetesimal Formation}
\item{Planetary Migration in Protoplanetary Disks}
\item{Formation of Giant Planets}
\item{Formation of Super-Earths}
\item{Debris Disks: Probing Planet Formation}
\item{Circumstellar Discs: What Will Be Next?}
\end{itemize}



\begin{acknowledgement}
The authors thank Takayuki Muto, Mariana Sanchez and Sid Gandhi for providing useful comments on this manuscript.


\end{acknowledgement}

\bibliographystyle{spbasicHBexo}  
\bibliography{dustevolution.bbl} 

\end{document}